\documentclass[reprint, prl, superscriptaddress, shortbibliography, floatfix, nobalance]{revtex4-1}

\usepackage{graphicx}
\usepackage{xcolor}
\usepackage{times}
\usepackage{physics}
\usepackage{siunitx}
\usepackage{chemformula}
\usepackage{float}

\usepackage{pdfpages}
\usepackage{etoolbox}

\makeatletter
\AtBeginDocument{\let\LS@rot\@undefined}
\makeatother

\begin{document} 

\author{T.C.~van Thiel}
\email{t.c.vanthiel@tudelft.nl}
\affiliation{Kavli Institute of Nanoscience, Delft University of Technology, Lorentzweg 1, 2628CJ Delft, Netherlands}

\author{W.~Brzezicki}
\affiliation{International Research Center MagTop, Institute of Physics, Polish Academy of Sciences, Warsaw, Poland}
\affiliation{M. Smoluchowski Institute of Physics, Jagiellonian University, prof. S. \L{}ojasiewicza 11, PL-30348 Krak\'ow, Poland }

\author{C.~Autieri}
\affiliation{International Research Center MagTop, Institute of Physics, Polish Academy of Sciences, Warsaw, Poland}

\author{J.R.~Hortensius}
\affiliation{Kavli Institute of Nanoscience, Delft University of Technology, Lorentzweg 1, 2628CJ Delft, Netherlands}

\author{D.~Afanasiev}
\affiliation{Kavli Institute of Nanoscience, Delft University of Technology, Lorentzweg 1, 2628CJ Delft, Netherlands}

\author{N.~Gauquelin}
\affiliation{Electron Microscopy for Materials Science (EMAT), University of Antwerp, Antwerp, Belgium}

\author{D.~Jannis}
\affiliation{Electron Microscopy for Materials Science (EMAT), University of Antwerp, Antwerp, Belgium}

\author{N.~Janssen}
\affiliation{Kavli Institute of Nanoscience, Delft University of Technology, Lorentzweg 1, 2628CJ Delft, Netherlands}

\author{D.~J.~Groenendijk}
\affiliation{Kavli Institute of Nanoscience, Delft University of Technology, Lorentzweg 1, 2628CJ Delft, Netherlands}

\author{J.~Fatermans}
\affiliation{Electron Microscopy for Materials Science (EMAT), University of Antwerp, Antwerp, Belgium}
\affiliation{Imec-Vision Lab, University of Antwerp, Wilrijk, Belgium}

\author{S.~van~Aert}
\affiliation{Electron Microscopy for Materials Science (EMAT), University of Antwerp, Antwerp, Belgium}

\author{J.~Verbeeck}
\affiliation{Electron Microscopy for Materials Science (EMAT), University of Antwerp, Antwerp, Belgium}

\author{M.~Cuoco}
\affiliation{Consiglio Nazionale delle Ricerche, CNR-SPIN, Italy}
\affiliation{Dipartimento di Fisica `E.R. Caianiello', Universit{\`a} degli Studi di Salerno, Fisciano Italy}

\author{A.D.~Caviglia}
\email{a.caviglia@tudelft.nl}
\affiliation{Kavli Institute of Nanoscience, Delft University of Technology, Lorentzweg 1, 2628CJ Delft, Netherlands}

%\begin{linenumbers}

\title{Coupling charge and topological reconstructions at polar oxide interfaces}

\begin{abstract}
In oxide heterostructures, different materials are integrated into a single artificial crystal, resulting in a breaking of inversion-symmetry across the heterointerfaces. A notable example is the interface between polar and non-polar materials, where valence discontinuities lead to otherwise inaccessible charge and spin states. This approach paved the way to the discovery of numerous unconventional properties absent in the bulk constituents. However, control of the geometric structure of the electronic wavefunctions in correlated oxides remains an open challenge. Here, we create heterostructures consisting of ultrathin \ch{SrRuO3}, an itinerant ferromagnet hosting momentum-space sources of Berry curvature, and \ch{LaAlO3}, a polar wide-bandgap insulator. Transmission electron microscopy reveals an atomically sharp \ch{LaO}/\ch{RuO2}/\ch{SrO} interface configuration, leading to excess charge being pinned near the \ch{LaAlO3}/\ch{SrRuO3} interface. We demonstrate through magneto-optical characterisation, theoretical calculations and transport measurements that the real-space charge reconstruction modifies the momentum-space Berry curvature in \ch{SrRuO3}, driving a reorganization of the topological charges in the band structure. Our results illustrate how the topological and magnetic features of oxides can be manipulated by engineering charge discontinuities at oxide interfaces.
\end{abstract}

\maketitle

%%%%%%%%%%%% Introduction %%%%%%%%%%%% 

\begin{figure*}[ht]
    \centering
    \includegraphics[width=1\linewidth]{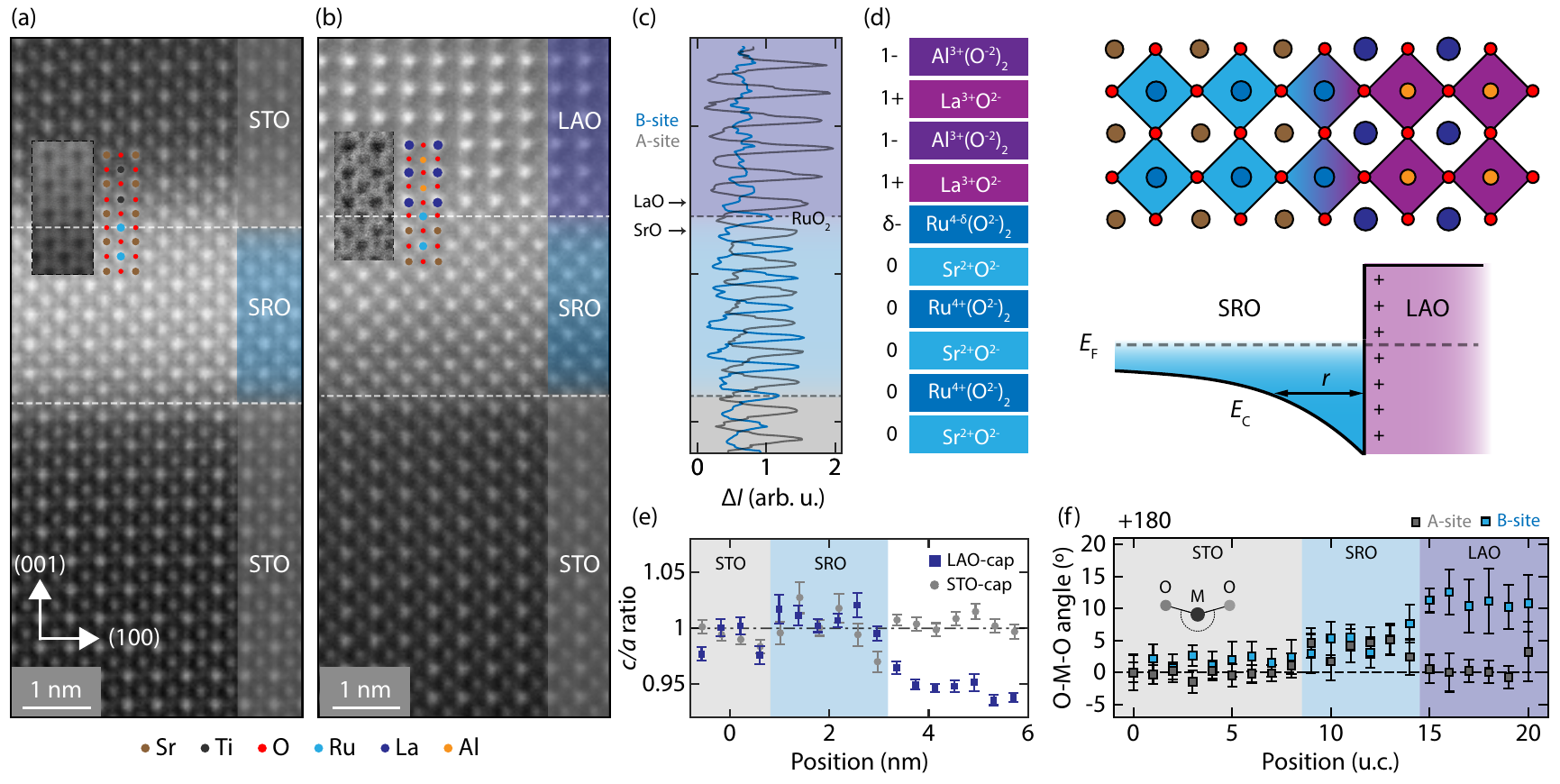}
    \caption{\textbf{Atomic characterization.} High-angle annular dark-field images of (a) STO/SRO/STO and (b) STO/SRO/LAO heterostructures with the insets showing bright-field images of the interface regions, (c) intensity profile along the growth-axis for the atomic \ch{A} (gray) and \ch{B}-sites (blue) of (b), (d) illustration of the charge frustration and the resulting profile of the chemical potential $E_C$ close to the Fermi energy $E_\text{F}$ at the \ch{LaO}/\ch{RuO2} interface, (e) $c/a$ ratio along the growth axis for (a) and (b), and (f) \ch{O}-\ch{M}-\ch{O} bond angles for the \ch{A} and \ch{B}-sites of the STO/SRO/LAO heterostructure, defined with respect to the STO substrate.}
    \label{Fig1Structure}
\end{figure*}

Recently, an increasing amount of attention has been focused on topological phases in condensed matter~\cite{wang2017topological}. Symmetry is a decisive element, as it can either be essential or detrimental for topological order~\cite{hasan2010colloquium, wen2017colloquium}. An iconic example is the quantum Hall effect, where the breaking of time-reversal symmetry is associated with a non-zero Chern number~\cite{haldane1988model}. A second example is the Weyl semimetal, which breaks either time-reversal symmetry, inversion symmetry or both~\cite{zyuzin2012weyl}. Transitions between different topological phases may be achieved through e.g. external electric or magnetic fields~\cite{zhang2013electric, chang2013experimental, qian2014quantum}, a change in chemical composition~\cite{bernevig2006quantum, konig2007quantum, hsieh2008topological, dziawa2012topological}, or application of pressure~\cite{xi2013signatures, liang2017pressure, ideue2019pressure}. While typically associated with an energy gap, such transitions are not limited to insulators and semimetals. They may also occur in strongly metallic systems~\cite{ying2019topological}, which are usually characterized by a high density of interacting electrons~\cite{pesin2010mott}. A candidate material is the itinerant ferromagnet \ch{SrRuO3} (SRO)~\cite{fang2003anomalous}, which over the past years has been the subject of intense research~\cite{matsuno2016interface, kan2018alternative, ohuchi2018electric, wang2018ferroelectrically, li2020reversible, van2020extraordinary, wang2020controllable, groenendijk2020berry,  kan2020electric, takiguchi2020quantum}. However, manipulating the properties of SRO-based heterostructures remains an experimental open challenge. Unlike insulators and semimetals, the high carrier density renders electrostatic gating, although possible~\cite{shimizu2014gate, mizuno2017electric, ohuchi2018electric, kan2020electric}, an inefficient method for manipulating the position of the Fermi level with respect to the momentum-space sources of Berry curvature. This calls for a different approach where the focus lies not on tuning the position of the Fermi level, but rather on changing the topological charges within the Brillouin zone i.e., inducing a topological transition. In this respect, oxide heterostructures provide an ideal platform due to the strong breaking of inversion-symmetry across the interfaces, especially between materials with different charge states~\cite{reyren2007superconducting, tsukazaki2007quantum, hwang2012emergent, skoropata2020interfacial}.\\
\noindent In this Letter, we demonstrate control of the momentum-space topological properties of ultrathin SRO, by creating a charge-frustrated interface. We synthesize \ch{RuO2}-terminated SRO ultrathin films and interface them with the polar wide-bandgap insulator \ch{LaAlO3} (LAO). The charge frustration leads to charge doping of SRO well beyond the capabilities of a conventional electrostatic gate therefore forming a pronounced profile of excess charge along the growth axis. We then demonstrate that in the ultrathin limit, this charge reconstruction modifies the momentum-space Berry curvature and leads to a full reversal of its sign for all temperatures below the magnetic transition, thereby controlling a topological transition in momentum-space. These results underline the potential of engineering charge discontinuities at oxide interfaces for inducing topological transitions in correlated matter.

%%%%%%%%%%%% Results %%%%%%%%%%%% 

\begin{figure}[ht]
    \centering
    \includegraphics[width=1\linewidth]{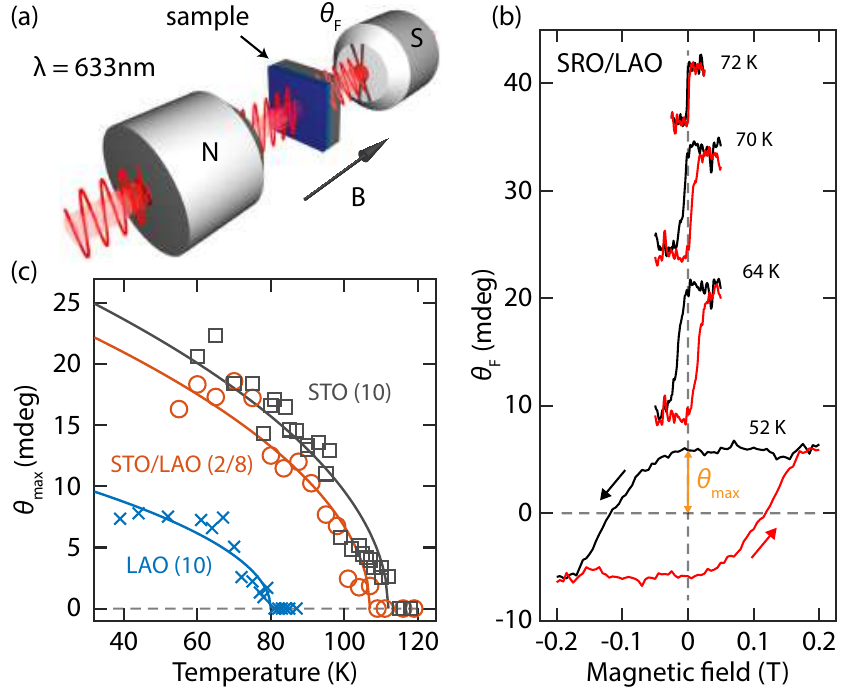}
    \caption{\textbf{Magneto-optical Faraday effect.} (a) Illustration of the experimental geometry for probing the Faraday rotation, (b) Faraday-rotation as a function of applied magnetic field for an SRO(5)/LAO(10) heterostructure for various temperatures and (c) the amplitude of the hysteresis loop $\theta_\text{max}$ as a function of applied field for SRO(5)/STO(10) (gray), SRO(5)/STO(2)/LAO(8) (orange) and SRO(5)/LAO(10) (blue) heterostructures. The solid lines represent Landau fits.}
    \label{Fig2Optics}
\end{figure}

In Fig.~\ref{Fig1Structure}a-c, we present scanning transmission electron microscopy data of a non-polar STO/SRO/STO and polar STO/SRO/LAO heterostructure, respectively. Due to both STO and SRO having the same \ch{Sr} A-site cation, the interface between these two \ch{ABO3} perovskites consists of \ch{BO2} layers (\ch{B} = \ch{Ti}, \ch{Ru}) separated by a shared \ch{SrO} plane. Consequently, both STO and SRO preserve the \ch{B^{4+}} valence state and the planar charges are zero on both sides of the interface. In contrast, we find that SRO and LAO are not separated by \ch{SrO}, but by a shared \ch{LaO} plane, indicating that the SRO film is \ch{RuO2}-terminated. This is a surprising observation since, due to the highly volatile nature of \ch{Ru_xO_y} species, the \ch{SrO}-termination has been argued to be more stable in oxidizing conditions~\cite{rijnders2004enhanced, koster2012structure}. The in-situ stabilization demonstrated here poses a substantial advantage over ex-situ approaches and is a promissing mechanism that invites further exploration~\cite{shin2005surface, lee2020atomic}. Irrespective of its origin, the observed \ch{LaO}/\ch{RuO2}/\ch{SrO} interface has important consequences for the \ch{Ru} charge state, with on the SRO side \ch{Sr^{2+}} requiring \ch{Ru^{4+}} and on the LAO side \ch{La^{3+}} requiring \ch{Ru^{3+}} for charge neutrality. The interface is effectively equivalent to the hybrid compound \ch{Sr_{0.5}La_{0.5}RuO3}. In a fully ionic picture, charge neutrality is then accomplished by a \ch{Ru^{3.5+}} charge state i.e., a $-0.5e$ excess charge at the interfacial layer (Fig.~\ref{Fig1Structure}d). Aside from the charge doping, the polarity in the LAO layers creates an attractive electric potential drawing charges towards the interface. Due to the abundance of free carriers in \ch{SrRuO3} ($n_e \sim 10^{22} \si{cm^{-3}}$), the corresponding electric field is screened over a length $r$, given approximately by the average distance between free carriers~\cite{slater1951simplification}. For our films, this yields $r \approx \qty({n_e})^{-\frac{1}{3}} = \SI{5}{\AA}$ or 1-2 crystal unit cells. Fig.~\ref{Fig1Structure}e shows the out-of-plane unit cell deformation along the growth axis. We find that for both heterostructures, the SRO film, as well as the STO and LAO overlayers, are coherently matched to the in-plane unit cell parameter of the substrate. The mismatch between the unit cell sizes is accommodated through a $c$-axis elongation in SRO and contraction in LAO (Fig.~\ref{Fig1Structure}e), indicating that the lattice structure is governed by the substrate and not by the capping layer~\footnote{This is further substantiated by a strong suppression of the octahedral tilts, yielding a tetragonal lattice symmetry, see Section I.B of the Supplementary Information.}. The absence of antiferrodistortive tilts indicates that the polar field in LAO must be compensated in another manner. In the well-known LAO/STO system, this is accomplished by a polar mode in the LAO layer, where the \ch{O}-\ch{Al}-\ch{O} bonds buckle in response to the internal electric field~\cite{huijben2009structure, gazquez2017competition}, a distortion which propagates into the top few unit cells of the STO substrate. Here, we observe a similar phenomenology i.e., a polar mode in the LAO layer that propagates into the top unit cell of the SRO layer (Fig.~\ref{Fig1Structure}f). 

\begin{figure}[ht]
    \centering
    \includegraphics[width=1\linewidth]{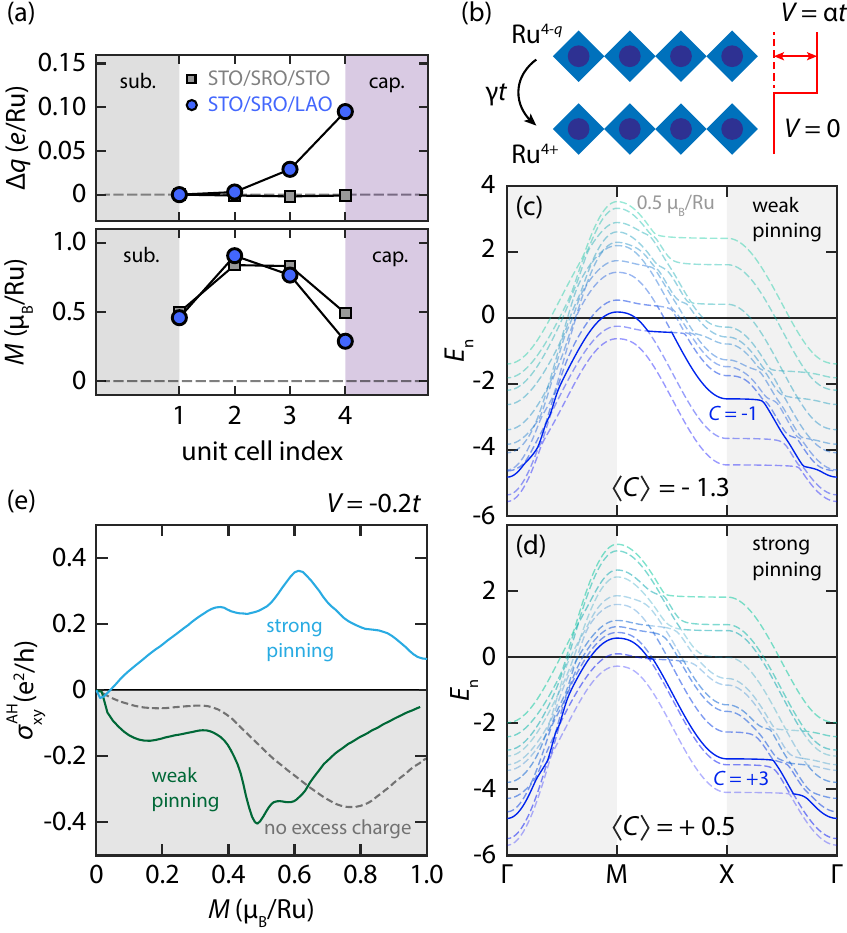}
    \caption{\textbf{Excess charge profile and topological reconstruction.} (a) Layer-resolved charge and magnetization profile for STO/SRO(4)/STO (gray) and STO/SRO/LAO (blue) calculated with DFT, (b) effective tight-binding model of two coupled SRO monolayers, with excess charge $q = -\SI{0.5}{e}$ in the top layer, as well as an on-site electrostatic potential $V$ and interlayer coupling $\gamma t$, (c-d) the 12 \ch{Ru} $t_{2g}$ bands for the effective model with (c) representing a weak pinning ($\gamma = 1$) and (d) a strong pinning ($\gamma = 0.5$) scenario, including an orbital Rashba SOC $\lambda_\text{R} = 0.04t$ correction and an on-site potential $V=-0.2t$ in the top layer. Panel (e) shows the calculated anomalous Hall conductivity $\sigma_{xy}^\text{AH}$, in which the gray dashed line represents the reference case with two $\ch{Ru}^{4+}$ layers ($q = 0$) and the colored solid lines represent the scenarios with excess charge $q = \SI{-0.5}{e}$ with either weak (green) or strong pinning (blue).}
    \label{Fig3Theory}
\end{figure}

We next investigate a second consequence anticipated for the excess charge accumulation i.e., a change in the spin state of the \ch{Ru} ions. In contrast to some of its magnetic $3d$ counterparts, the crystal-field splitting in SRO ($\sim\SI{3}{eV}$) is larger than both the Hund's interaction ($\sim \SI{0.3}{eV}$) and the Coulomb repulsion ($\sim\SI{2}{eV}$), owing to the spatially larger $4d$ orbitals~\cite{lee2001optical, laad2001origin, dang2015electronic}. As a result, the four $d$ electrons in SRO reside in the $t_{2g}$ orbitals, producing a ($4d^4$) $S=1$ spin state. The additional charge transferred from the LAO layer also occupies the $t_{2g}$ manifold, reducing the average spin state to $3/4 < S_\text{avg} < 1$ due to the spin flip. Such a reduction of the spin moment directly manifests in the value of the saturation magnetization $M_\text{s}$ and the Curie temperature $T_\text{C}$, which both scale with $S(S+1)$. In magnetic systems, both $T_\text{C}$ and $M_\text{s}$ can thus serve as indicators of a charge reconstruction and are expected to be lowered when electronic charge is transferred to SRO. We therefore proceed to investigate the spontaneous magnetization as a function of temperature, by means of the magneto-optical Faraday effect (Figs.~\ref{Fig2Optics}a-b). The polarization rotation $\theta_\text{F}$ due to the Faraday effect is linearly proportional to the out-of-plane component of the magnetization $M_z$. In Fig.~\ref{Fig2Optics}c, we show the Faraday rotation as a function of temperature for SRO films with various capping layers. To compare the different $M_\text{s}$ and $T_\text{C}$, we fit the data to $\theta = \theta_{T=0} \abs{T-T_\text{C}}^{1/2}$, treating $\theta$ as the magnetization $M_\text{s}$. According to expectation, we find a clear suppression of both $T_\text{C}$ and $\theta_F$ for the SRO(5)/LAO(10) sample, as compared to SRO(5)/STO(10). To verify that this is an interface-driven effect, we also investigated SRO(5)/STO(2)/LAO(8), which is structurally similar to SRO(5)/LAO(10), but has two layers of STO that shield SRO from the valence discontinuity. As expected, both $M_\text{s}$ and $T_\text{C}$ are significantly larger compared to the LAO-capped sample and nearly identical to the STO-capped sample, further supporting that the charge and magnetic reconstruction are driven by the charge frustration at the \ch{LaO}/\ch{RuO2} interface.

Having established an interface-driven charge and spin reconstruction, we turn to the question of how this affects the momentum-space Berry curvature and the anomalous Hall effect (AHE). Aside from the altered magnetization, the charge frustration introduces two elements; (i) a shift in the chemical potential due to charge doping and (ii) breaking of inversion-symmetry due to the electric field along the growth axis. To determine their impact, we first address the question of how far the field penetrates into the SRO film. The top panel in Fig.~\ref{Fig3Theory}a shows the DFT-calculated charge profile across the heterostructure, using the STO substrate as a baseline value. As expected, charge doping is absent for the STO-capped heterostructure, yielding a symmetric charge profile. For the LAO-capped sample however, we find, in accordance with the previous estimate for the screening length $r$, a doping of $\SI{\sim -0.1}{e}$ and $\SI{\sim -0.04}{e}$ for the two unit cells closest to the SRO/LAO interface ($\SI{\sim e21}{cm^{-3}}$), leading to a strongly asymmetric charge profile~\footnote{Note that the charges do not add up to $\SI{-0.5}{e}$ due to the covalent bonding between ions considered in DFT i.e., a considerable portion of the additional charge resides in the interstitial space and oxygen ligands.}. The impact on the magnetization is immediately clear from the bottom panel of Fig.~\ref{Fig3Theory}a, which, in agreement with the magneto-optical characterization, shows that the two charge-reconstructed unit cells have a lower magnetization compared to the STO/SRO/STO reference case. These results provide a clear picture; the SRO film experiences an electronic and magnetic reconstruction that persists 2 u.c. from the interface and causes a strong inversion-symmetry breaking. 

To determine the effect of the reconstruction on the momentum-space Berry curvature, we introduce an effective tight-binding model of an SRO bilayer with interlayer coupling $\gamma t$, where $t$ is the nearest-neighbour hopping energy (Fig.~\ref{Fig3Theory}b). The charge-frustration and symmetry-breaking is simulated by including an additional charge $-0.5e$ in the top layer, an attractive electrostatic potential $V=-0.2t$ and a small orbital Rashba correction $\lambda_R = 0.04t$. The parameter $\gamma$ represents the tendency of the excess charge being pinned to the top layer i.e., the screening length $r$ defined in Fig.~\ref{Fig1Structure}d. From a band structure perspective, this translates to a steeper bending of the chemical potential $E_C$ near the interface. We consider two scenarios; weak and strong charge pinning, or high and low $\gamma$, respectively. Figs.~\ref{Fig3Theory}c-d show the dispersion relations of the twelve \ch{Ru} $t_{2g}$ bands for the two scenarios, at a representative value of the magnetization. The kinks that can be observed at the band anticrossings represent momentum-space (anti-)vortices of the Berry connection, acting as either positive or negative charges of Berry curvature. We highlight one band as an example, whose Chern number transitions from $C = -1$ to $C = +3$ between the weak and strong pinning scenarios. In the dispersion, this manifests as a change in the position and character of the band (anti-)crossings. Overall, we find a substantial evolution of the topological charges between the two scenarios. One can approximate the total Berry curvature by the averaged Chern number $\expval{C}$, which is calculated by summing the Chern numbers of the individual bands, weighted by their occupation. We find a transition of $\expval{C} = -1.3$ to $\expval{C} = +0.5$, between the weak and strong pinning scenarios, respectively. Concurrently, the filling factors between the two scenarios remain virtually unchanged. In fact, it can be shown that for any linearly decreasing profile of the filling factors with energy, the sign of the total Berry curvature is purely determined by the sum of the Chern numbers associated with the indirect gaps of the twelve $t_{2g}$ bands~\cite{Supplementary}. This demonstrates that the sign change is driven by a topological transition in momentum-space and is not due to a change in band occupation. To further demonstrate the robustness of this result, we directly calculate the anomalous Hall conductivity for a wide range of values for the magnetization (Fig.~\ref{Fig3Theory}e). In agreement with the topological charge reconstruction, we find a transition from a fully negative to a fully positive AHE for nearly all magnetization values. These results unambiguously identify the charge pinning, and the resulting inversion-symmetry breaking, as the dominant effect in reconstructing the momentum-space topological charges and Berry curvature.

\begin{figure}[ht]
    \centering
    \includegraphics[width=1\linewidth]{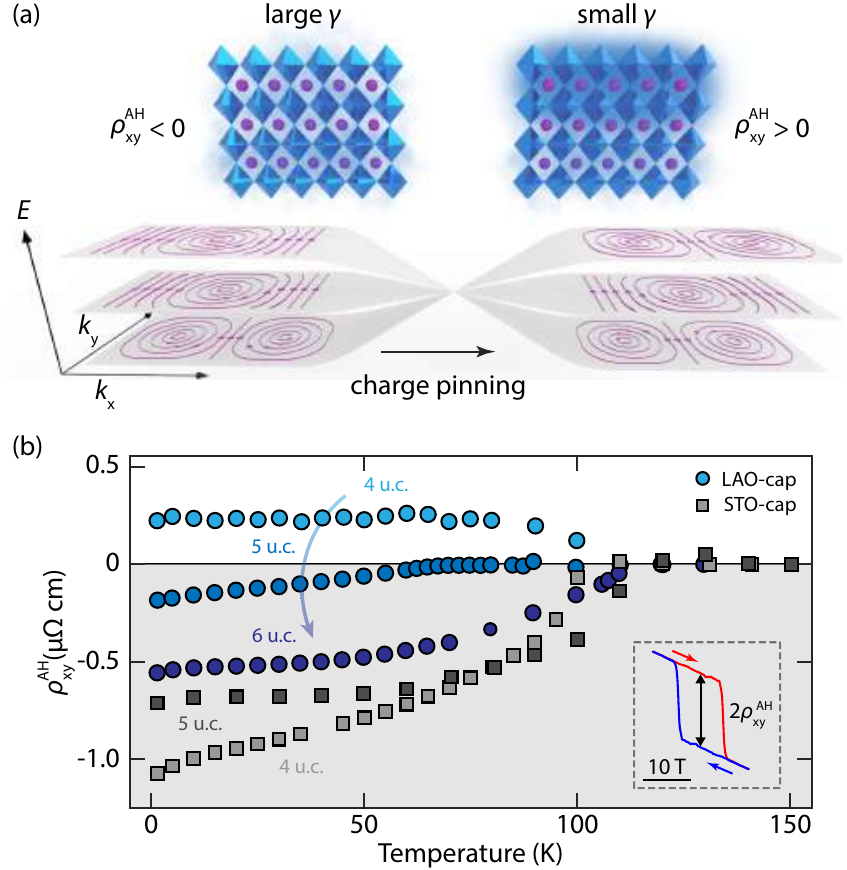}
    \caption{\textbf{Anomalous Hall effect.} (a) Illustration representing the evolution of the momentum-space topological charges. Upon increasing the charge pinning, the system moves through a Weyl point in the synthetic space spanned by $k_x$, $k_y$ and the charge pinning parameter $\gamma$, (b) shows the measured anomalous Hall resistivity $\rho_{xy}^\text{AH}$ for SRO films of varying thickness capped by both STO and LAO as a function of temperature. The inset in (b) shows an example of the magnetic-field dependence of the AHE, from which the amplitude $\rho_{xy}^\text{AH}$ is extracted.}
    \label{Fig4Transport}
\end{figure}

To illustrate the topological reconstruction, we consider a three-dimensional parameter space, spanned by the Bloch momentum coordinates $k_x$ and $k_y$ and the charge pinning parameter $\gamma$ as the third dimension. The reconstruction can then be understood as the system moving through a Weyl point, where the bands experience a closing and reopening of an energy gap upon increasing the charge pinning parameter $\gamma$. The evolution changes both the number and sign of the topological charges in the two-dimensional Brillouin zone. We have visualized this concept in Fig.~\ref{Fig4Transport}a, where the sources of Berry curvature are represented by chiral vortex-like objects of the Berry connection, which change both in winding direction and in number across the transition. In a transport experiment, this translates to an inversion of the sign of the emergent field and consequently the AHE. Fig.~\ref{Fig4Transport}b shows the AHE amplitude, as a function of temperature for films of various thicknesses $m$ capped by STO and LAO. For the thinnest films ($m = 4$), we find a positive AHE for the LAO-capped sample for all temperatures below the $T_\text{C}$. In contrast, it is negative for the 4 u.c. STO-capped sample. These two heterostructures represent the left and right scenarios in Fig.~\ref{Fig4Transport}a and the sign inversion of the AHE can be understood as the system experiencing a topological reconstruction, driven by the onset of charge pinning at the polar interface. As $m$ is increased for the LAO-capped samples, we find a transition to a more negative behavior of the AHE in temperature, which can be understood as the contribution of the charge-reconstructed layers becoming increasingly diluted as the total film thickness increases, effectively diminishing the impact of the interface inversion symmetry-breaking on the anomalous Hall response. Accordingly, one expects both heterostructures to converge to the same state as the number of layers is increased. Indeed, for increasing layer thickness, both heterostructures tend towards the same behavior of the AHE i.e., a negative at all temperatures.  

%%%%%%%%%%%% Discussion & Conclusion %%%%%%%%%%%% 

While electric field penetration in bulk metals can be safely ignored, its importance in the ultrathin limit cannot be neglected. The key element is the extremely short penetration depth of the electrostatic potential in metals, which can cause a strong inversion-symmetry breaking in the near-interface region. The resulting electronic and magnetic reconstructions can have a decisive effect on the momentum-space topological properties of correlated systems, including, but not limited to \ch{SrRuO3}. The charge frustration arising from the interface with \ch{LaAlO3} provides a unique opportunity for studying the effect of symmetry-breaking on its momentum-space topology. Due to the insulating nature of \ch{LaAlO3}, there is neither mixing of states at the Fermi energy nor interface-driven spin canting, as has been reported in e.g. the \ch{SrRuO3}/\ch{SrIrO3} interface, which has been the topic of multiple studies in recent years~\cite{matsuno2016interface, pang2017spin, ohuchi2018electric, groenendijk2020berry, zeng2020emergent}. In this sense, the system considered here offers a pleasing simplicity and a more direct approach towards controlling the topology in ultrathin \ch{SrRuO3} and potentially other correlated metals. Our results are also of relevance to the scenario of uncapped \ch{SrRuO3} films~\cite{gu2019interfacial}, where dangling bonds at the surface can manifest as a charged electrostatic boundary condition~\cite{lee2020atomic}, albeit complicated by the unavoidable interaction with adsorbed ambient chemical species.\\ In conclusion, we have demonstrated how a valence charge discontinuity induces both a magnetic and topological reconstruction in ultrathin films of the itinerant ferromagnet \ch{SrRuO3}. We identify the pinning of the excess charge donated by the polar \ch{LaAlO3} overlayer and the resulting inversion-symmetry breaking to be the dominant effect in altering the band topology and momentum-space Berry curvature, leading to a full inversion of the sign of the emergent magnetic field. These results demonstrate how engineering charge discontinuities can be utilized to control the topological properties in oxide heterostructures and establish the potential of interface design towards the manipulation of the geometric structure of wavefunctions in correlated matter.\\

%%%%%%%%%%%% Acknowledgements %%%%%%%%%%%% 

The authors thank E.~Lesne, M.~Lee, H.~Barakov, M.~Matthiesen and U.~Filippozzi for discussions. The authors are grateful to E.J.S. van Thiel for producing the illustration in Fig.~\ref{Fig4Transport}a. This work was supported by the European Research Council under the European Unions Horizon 2020 programme/ERC Grant agreements No. [677458], [770887] and No. [731473] (Quantox of QuantERA ERA-NET Cofund in Quantum Technologies) and by the Netherlands Organisation for Scientific Research (NWO/OCW) as part of the Frontiers of Nanoscience (NanoFront) and VIDI program. The authors acknowledge funding from the European Union’s Horizon 2020 research and innovation programme under grant agreement No. [823717] - ESTEEM3. N. G., J. V., and S. V. A. acknowledge funding from the University of Antwerp through the Concerted Research Actions (GOA) project Solarpaint and the TOP project. C. A. and W. B. are supported by the Foundation for Polish Science through the International Research Agendas program co-financed by the European Union within the Smart Growth Operational Programme. C. A. acknowledges access to the computing facilities of the Interdisciplinary Center of Modeling at the University of Warsaw, Grant No.~G73-23 and G75-10. W.B. acknowledges support from the Narodowe Centrum Nauk (NCN, National Science Centre, Poland) Project No. 2019/34/E/ST3/00404.\\ 

\noindent Additional data, as well as a detailed description of the sample synthesis, TEM characterization, magnetic characterization and theoretical analysis are discussed in the Supplementary Information~\cite{Supplementary}.

\bibliography{References.bib}
\newpage


\section*{Supplementary material}

# Supplementary Information for:

# Coupling charge and topological reconstructions at polar oxide interfaces

T.C. van Thiel,[1, *] W. Brzezicki,[2, 3] C. Autieri,[2] J.R. Hortensius,[1]

D. Afanasiev,[1] N. Gauquelin,[4] D. Jannis,[4] N. Janssen,[1] D. J. Groenendijk,[1]

J. Fatermans,[4, 5] S. van Aert,[4] J. Verbeeck,[4] M. Cuoco,[6, 7] and A.D. Caviglia[1, †]

[1]*Kavli Institute of Nanoscience, Delft University of Technology, Lorentzweg 1, 2628CJ Delft, Netherlands*
[2]*International Research Center MagTop, Institute of Physics,*
*Polish Academy of Sciences, Warsaw, Poland*
[3]*M. Smoluchowski Institute of Physics, Jagiellonian University,*
*prof. S. Łojasiewicza 11, PL-30348 Kraków, Poland*
[4]*Electron Microscopy for Materials Science (EMAT), University of Antwerp, Antwerp, Belgium*
[5]*Imec-Vision Lab, University of Antwerp, Wilrijk, Belgium*
[6]*Consiglio Nazionale delle Ricerche, CNR-SPIN, Italy*
[7]*Dipartimento di Fisica 'E.R. Caianiello', Università degli Studi di Salerno, Fisciano Italy*

## CONTENTS

---

* t.c.vanthiel@tudelft.nl

† a.caviglia@tudelft.nl

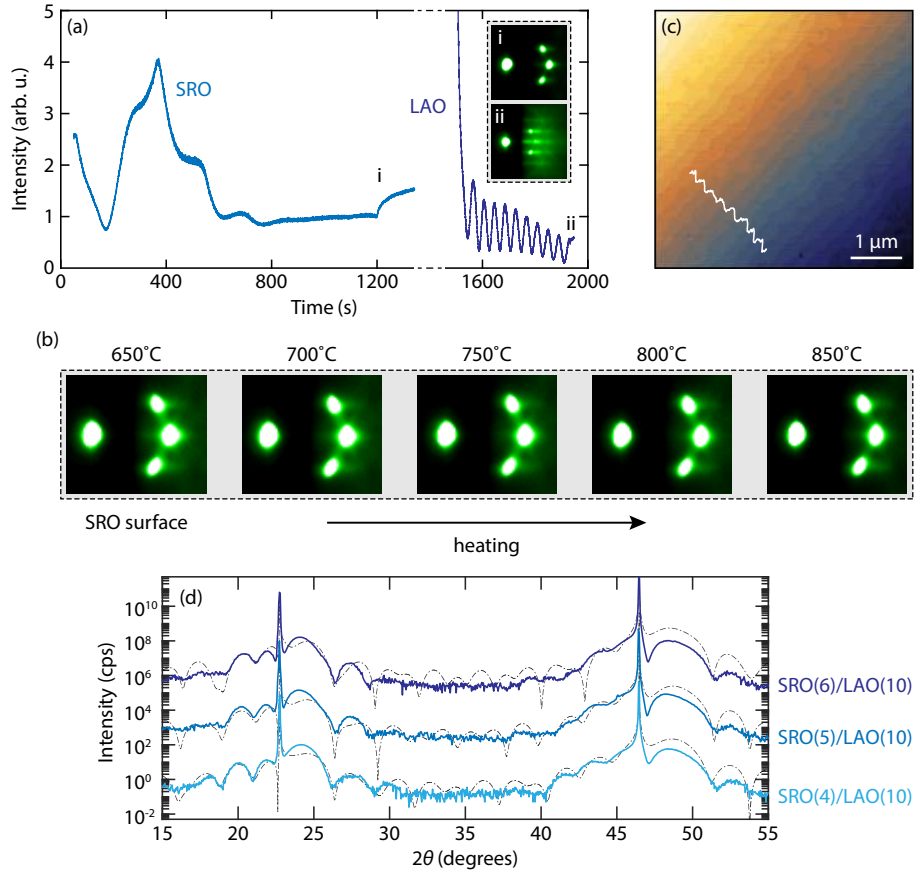

FIG S1. Surface characterization. (a) intensity of the first-order RHEED spot during the growth of an SRO/LAO heterostructure. The inset shows the RHEED pattern after the growth of (i) the SRO and (ii) the LAO layers. (b) RHEED pattern of the SRO surface during heating ($p_{O_2} = 6 \times 10^{-5}$ mbar), prior to the deposition of LAO and (c) an AFM topographic image of the LAO surface. The inset shows the height profile. Panel (e) shows $\theta$-$2\theta$ X-ray diffraction scans with the solid lines representing the data and the black dashed line simulations [1]. The curves are shifted vertically for visual clarity.

# I. STRUCTURAL, MAGNETIC AND TRANSPORT CHARACTERIZATION

## A. Synthesis

All samples were grown on (001) $TiO_2$-terminated $SrTiO_3$ (STO) substrates by pulsed-laser deposition (PLD), using a 248 nm KrF excimer laser at a 1 Hz repetition rate. The laser fluences were $1.2$ Jcm$^{-2}$ for SRO and $1.0$ Jcm$^{-2}$ for the LAO and STO capping layers. STO substrates were mechanically clamped

onto a heater stage. The substrates were heated by an infrared diode laser and the temperature was measured with a pyrometer. SRO layers were deposited at 600 °C in an oxygen partial pressure of 0.1 mbar. The growth was monitored with in-situ RHEED, recording the first-order diffraction spot (see Fig. S1a). The first oscillation corresponds to a change in surface termination of the substrate from $TiO_2$ to SrO [2]. Subsequent oscillations represent depositions of full unit cells (consisting of $RuO_2$ and SrO planes). Starting from a SrO-terminated substrate, this nominally leads to a SrO-terminated film. The deposition is stopped at an intensity maximum, after which a strong recovery is observed, indicative of a high surface mobility of atomic species in the top layer. The final RHEED pattern of the SRO surface is shown in the inset and exhibits intense spots, implying an atomically smooth surface. Next, the oxygen partial pressure is reduced to $6 \times 10^{-5}$ mbar and the sample is further heated to 800 °C at a rate of 10 °C/min. To verify that the SRO surface does not degrade, the RHEED patterns were monitored during heating (see Fig. S1b). Evidently, the surface quality is maintained up to high temperatures ($> 850$ °C). Considering the high-surface mobility of the adatoms and the tendency of SrO to desorb at high temperatures and low partial pressures [3], we attribute the $RuO_2$-termination of the SRO/LAO heterostructures to the desorption of SrO. Another possible contribution is the larger electronegativity of Ru (2.0) and La (1.1) compared to Al (1.5) and Sr (1.0) [4], suggesting a chemically more stable (oxygen-mediated) connection between Ru and La, compared to Sr and Al. Once the sample temperature has reached the target value of 800 °C, a 10 u.c. LAO layer is deposited in a layer-by-layer growth mode (Fig. S1a). An AFM topographic image of the LAO surface is shown in Fig. S1c, which shows the heterostructure adapts the step-and-terrace morphology of the substrate. Finally, to refill oxygen vacancies that may have formed during the growth (which have been shown to play a role in the magnetization [5]), all samples are annealed at 550 °C in 300 mbar $O_2$ for 1h and subsequently cooled to room temperature at 20 °C/min in the same pressure. STO capping layers were grown in both high pressure, low temperature (600 °C, 0.1 mbar $O_2$), as well as low pressure, high temperature conditions (750 °C, $1 \times 10^{-4}$ mbar $O_2$). Regardless of the STO growth conditions, the Curie temperature of the SRO film is always significantly higher compared to LAO-capped samples and the AHE is always negative, except in the direct vicinity of the Curie point. Fig. S1d shows a $\theta$-$2\theta$ X-ray diffraction measurements (solid line), as well as simulations (dashed lines) [1]. The best simulations were obtained by imposing the designated number of SRO unit cells with $c = 3.98$ Å (compressive strain) and 10 u.c. LAO with $c = 3.72$ Å (tensile strain) for the LAO capping layer. Deviations from the simulation may be attributed to unit cell compression/elongation at the interfaces, as shown in Fig. 1 of the main text. The observed finite-size oscillations are due to long-range crystal order, indicative of a high sample quality on a macroscopic scale.

## B. Transmission Electron Microscopy

The samples were prepared in a vacuum transfer box and analyzed in a Gatan vacuum transfer holder to avoid any influence of air on the film. Atomic scale characterization of the lattice structure was performed on a $C_s$ corrected FEI Titan STEM microscope operated at 300 kV. The convergence angle of the incoming probe was 21 mrad and the inner and outer angle of the HAADF and ABF detector were

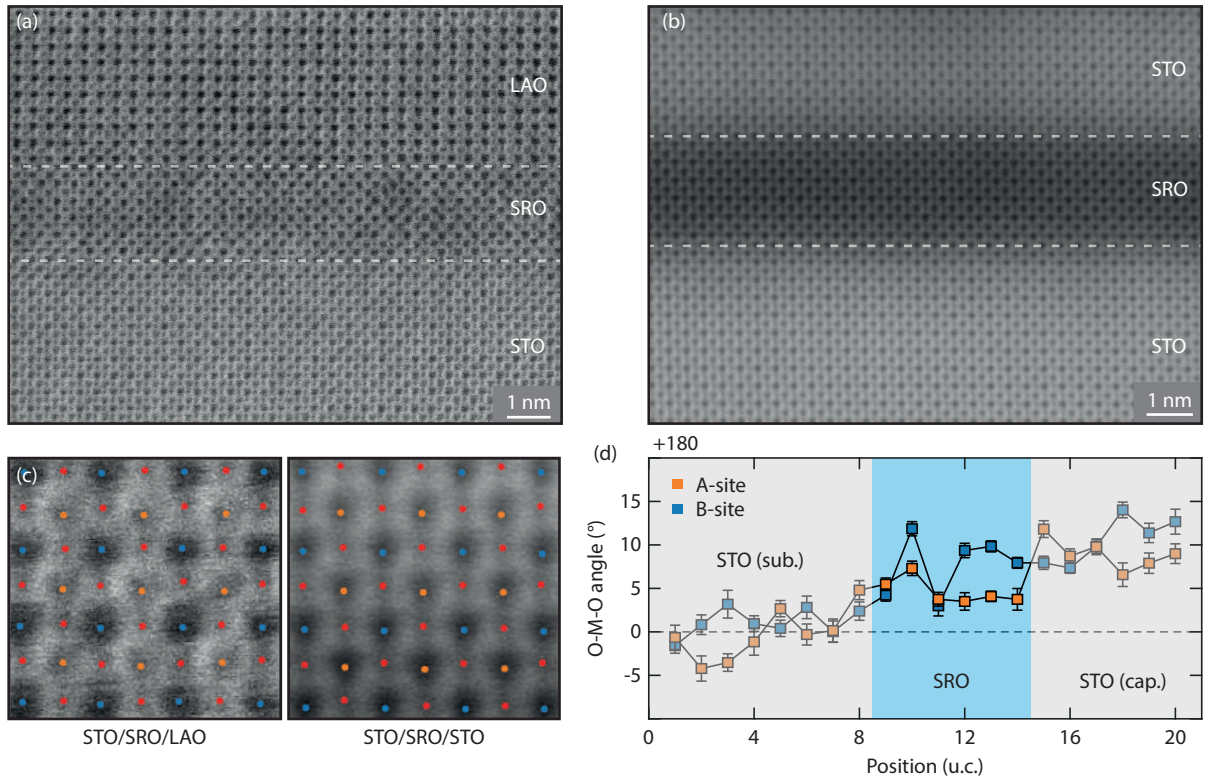

FIG S2. ABF imaging. Annular bright-field images of the (a) STO/SRO/LAO and (b) STO/SRO/STO heterostructures. Panel (c) illustrates the fitted atomic positions for the (red) oxygen anion, as well as the (orange) A- and (blue) B-site cations across the heterostructures, defined with respect to the STO substrate. Panel (d) shows the bond angle profile for the STO/SRO/STO heterostructure.

44-190 mrad and 8-17 mrad, respectively. The HAADF and ABF were acquired simultaneously as explained in ref. [6]. The HAADF mode (High-Angle Annular Dark-Field) is used to find the atom position of the heavier atoms whereas the ABF mode (Annular Bright-Field) has the ability to detect simultaneously the lighter and heavier atoms, which is used to obtain the atomic positions of the oxygen ions [7]. A first approximation of the atomic positions in both ADF and ABF is determined using the peak finder algorithm of the open-source software StatSTEM [8]. The refinement of the atom positions is performed by modeling both images as a superposition of 2D Gaussian peaks as described in ref.[9].

In Fig. S2, we show the ABF images of an STO/SRO/LAO and STO/SRO/STO heterostructure. In agreement with the HAADF images discussed in the main text, the vertically adjacent positions of La ($Z = 57$) and Ru ($Z = 44$) atoms across the SRO/LAO interface are indicative of a LaO/RuO$_2$ configuration. Fig. S2c illustrates the fitted positions for the oxygen atoms, as well as the A- and B-site cations, which are used to determine the profile of the O$-$M$-$O bond angle (M=A, B) along the growth axis. The

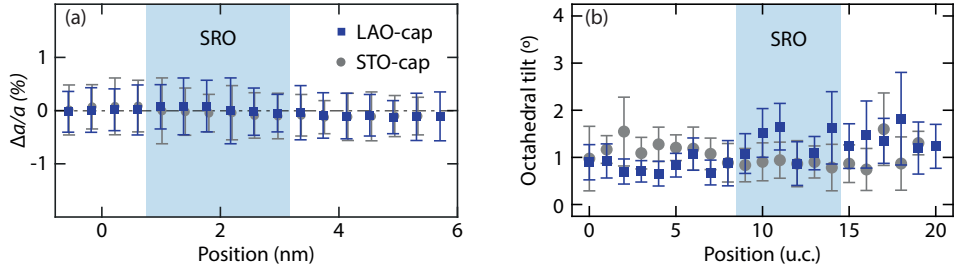

FIG S3. Strain and tilts. (a) In-plane unit cell deformation and (b) octahedral tilts along the growth axis for STO- and LAO-capped SRO films.

results for the STO/SRO/LAO heterostructure have been discussed in the main text, where buckling of the O−Al−O bond in the LAO capping layer was found to propagate into the top u.c. of the SRO film. We note that this effect was absent for the O−La−O bond. The angles for the STO/SRO/STO heterostructure are presented in Fig. S2d. In contrast to the STO/SRO/LAO heterostructure, we observe buckling of both the O−A−O and O−B−O bonds that becomes more pronounced towards the sample surface. We also find somewhat larger buckling angles in the SRO film, which may be due to the larger angles observed in the STO capping layer. This in turn may be attributed to a degree of off-stoichiometry promoting a polar mode, a common occurrence in STO thin films [10]. The absence of additional contributions in the AHE for the STO-capped samples indicates that structural features such as buckling are in all likelihood not the dominant source of the observed Berry curvature reconstructions.

Fig. S3 shows the in-plane unit cell deformation and octahedral tilts along the growth axis for the two heterostructures. We find that the lattice parameters of the SRO film and the LAO- and STO-capping layers are coherently matched to the STO substrate ($a = 3.905$ Å). In agreement with the XRD characterization shown in Fig. S1d. The unit cell size mismatch is then accomodated through compressive and tensile strain in the SRO and LAO layers, respectively. We also find a strong suppression of the octahedral tilts for both samples, yielding a tetragonal (almost cubic) lattice structure.

### C. Magneto-optical characterization

Magnetic characterization of the heterostructures was performed using the magneto-optical Faraday effect (see Fig. S4a), which is proportional to the out-of-plane component of the sample magnetization $M_z$ via

$$\theta_\mathrm{F} = V M_z t, \tag{1}$$

where $t$ is the sample thickness and $V$ the Verdet constant. Representative samples were grown on double-side polished STO substrates to allow for transmission measurements. The rotation of the polarization plane of linearly polarized laser light ($\lambda = 633$ nm, $h\nu = 2.0$ eV), was measured at normal

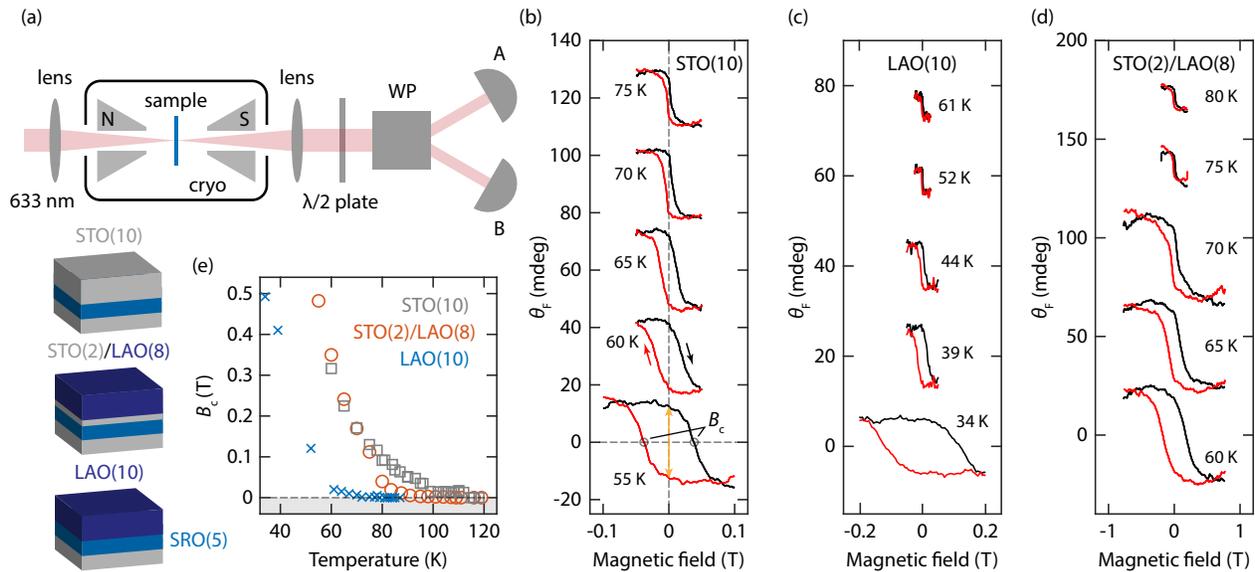

FIG S4. **Magneto-optical Faraday effect.** (a, top) Illustration of the experimental setup for probing the magneto-optical Faraday effect and (bottom) an illustration of the three different samples measured, (b-d) Faraday rotation $\theta_F$ as a function of magnetic field in 5 u.c. SRO films for different capping layers (b) 10 u.c. STO, (c) 10 u.c. LAO and (d) 2 u.c. STO followed by 8 u.c. LAO. The curves are shifted vertically for visual clarity. Panel (e) shows the coercive field $B_c$ as a function of temperature for the same samples of panels (b-d).

incidence as a function of applied (out-of-plane) magnetic field and temperature. The changes in the polarization of the transmitted light were measured using a Wollaston prism, which splits the beam into two orthogonally polarized components. The difference in intensity of the two beam components resulting from the magneto-optical Faraday effect is proportional to the polarization rotation ($\theta_F$) and was measured with a pair of balanced photodiodes. The small thicknesses of the samples and the (transmission) geometry of the experiment ensure that the entire SRO film is homogeneously probed and interference effects are avoided. The Faraday rotation of three heterostructures with different capping layers is shown in Fig. S4b-d. As discussed in the main text, the height of the hysteresis loop (indicated by the orange arrow) is directly proportional to the out-of-plane component of the sample magnetization. As discussed in the main text, we find a suppression of the Faraday rotation for the LAO(10) capping layer compared to the STO(2)/LAO(10) and STO(10) capping layers, which can be attributed to the altered valence and spin state of Ru at the RuO$_2$/LaO interface. We also find a suppression of the coercive field (Fig. S4e), indicating a weakened magnetic anisotropy. In line with the Stoner-Wohlfarth model for ferromagnetism, this further points to a reduced magnetization.

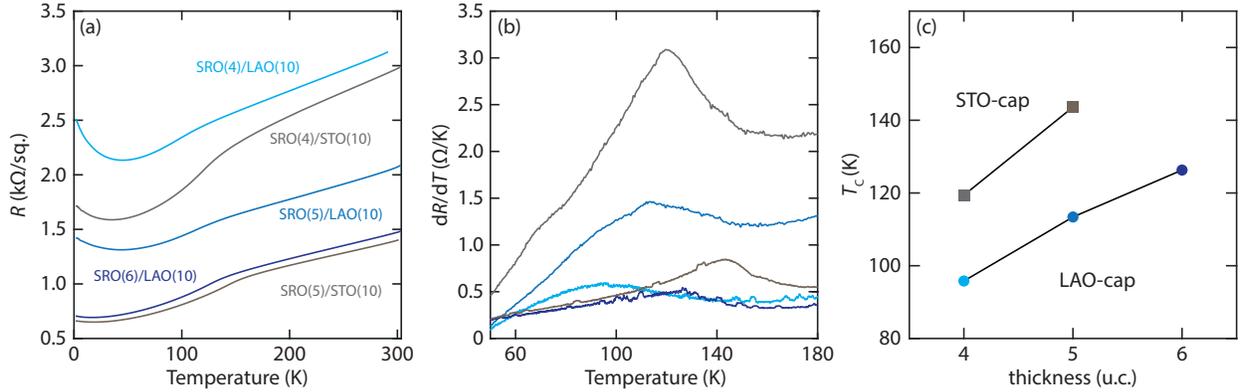

FIG S5. Transport characterization. (a) Sheet resistance as a function of temperature, (b) the corresponding derivatives and (c) the Curie temperatures for various STO- and LAO-capped SRO ultrathin films.

## D. Transport

In this section, we show extended transport data of the films discussed in the main text. We present data of both LAO- and STO-capped samples, with thicknesses between 4-6 u.c. All measurements are performed in large (mm-sized) Hall bar geometries, averaging out any spatial inhomogeneities in the transport properties. Longitudinal and transverse resistances are obtained by sourcing a low frequency ($\sim 17\,\mathrm{Hz}$) $10\,\mu\mathrm{A}$ current and measuring the resulting voltage drops with a lock-in amplifier.

In Figure S5a, we show the sheet resistance as a function of temperature for various STO- and LAO-capped SRO ultrathin films. According to expectation, we observe a lowering of the sheet resistance with an increasing amount of SRO layers. For the thinnest films, slight upturns at low temperature can be observed, indicative of disorder-induced carrier localization resulting from the close proximity to the thickness-driven metal-insulator transition [11]. At the Curie temperature $T_{\mathrm{C}}$, $R(T)$ exhibits an inflection point, caused by short-range spin fluctuations across the magnetic phase transition [12]. The inflection point corresponds to a peak in the $\mathrm{d}R/\mathrm{d}T$ (see Fig. S5b), the center of which we use to identify $T_{\mathrm{C}}$ (see Fig. S5c). We find (1) a suppressed Curie temperature in LAO-capped samples and (2) a requirement of 2 additional unit cells in LAO-capped samples to recover the $T_{\mathrm{C}}$ of the STO-capped samples. As shown in Fig. 4 of the main text, this observation also extends to the AHE.

In Figure S6, we show the magnetic-field-dependent Hall effect. Note the opposite orientation of the loops between the SRO(4)/LAO and SRO(4)/STO sample for all temperatures displayed (Fig.S6a-b), a clear indication of an opposite sign of the momentum-space Berry curvature. The peaks seemingly superposed on the hysteresis loop for the LAO-capped sample have over the past years been a source of debate, with them either being attributed due to real-space magnetic skyrmions (see e.g. refs. [13, 14]), or due to a spatially inhomogeneous profile of the magnetization (e.g. refs. [15, 16]). They only occur in the vicinity of the coercive field (i.e., during the magnetization reversal) and do not affect the anomalous

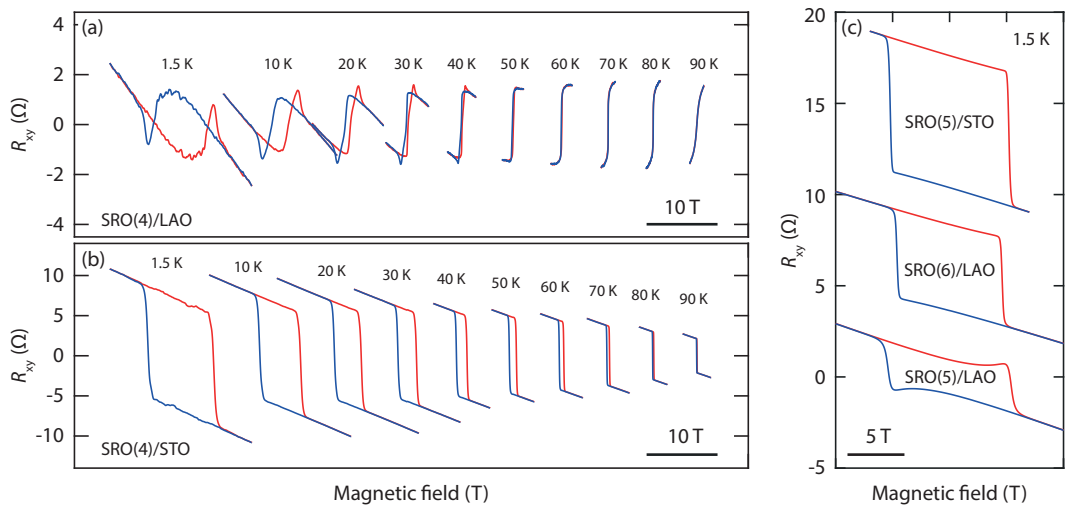

FIG S6. Hall effect as a function of magnetic field and for various temperatures for (a) an SRO(4)/LAO, (b) an SRO(4)/STO heterostructure. Panel (c) compares the magnetic-field dependent Hall effect at 1.5 K for SRO(5)/LAO, SRO(6)/LAO and SRO(5)/STO.

amplitude in the fully magnetized state. While interesting in their own right, the peaks are not relevant to the discussion on the momentum-space sources of Berry curvature considered here. We also note that they are absent in STO-capped samples and vanish in the LAO-capped samples as the SRO layer thickness is increased (see Fig. S6c).

In Figure S7, we show the Hall effect in the vicinity of the Curie point for the SRO(4)/LAO(10) and SRO(4)/STO(10) samples. A state is considered ferromagnetic if a finite remanence is present. For both heterostructures we find that as the temperature is increased towards $T_C$ the hysteresis tends to zero, but a remanence remains observable as a vertical 'jump' at low field. When $T_C$ is exceeded, the system enters a phase characterized by fluctuating residual moments [17–19], which are readily aligned by an external magnetic field. In the Hall effect, this manifests as the presence of an appreciable slope at zero magnetic field, yet the absence of a remanence. For the LAO-capped sample, we observe a deviation from a perfect vertical slope in the vicinity of 100 K. For the STO-capped samples the same phenomenology is observed, but shifted up in temperature by 20-30 K. These observations are in agreement with the $R$-$T$ curves shown in Figure S5, as well as the magneto-optical characterizations shown in the main text, although we find that the latter slightly underestimates the $T_C$ for all samples. This discrepancy can be attributed to the detection limit of our magneto-optical setup. Just below $T_C$, the magnetization is small and might fall below the magneto-optical detection limit, whereas resistivity measurements are surprisingly sensitive to the ferromagnetic transition due to short-range spin-fluctuations across the transition, manifesting as a pronounced peak in the $dR/dT$ (see e.g. ref. [18]). In addition, the magnetization of SRO films has a somewhat stronger in-plane component near $T_C$ [20], which is not probed in the Faraday configuration.

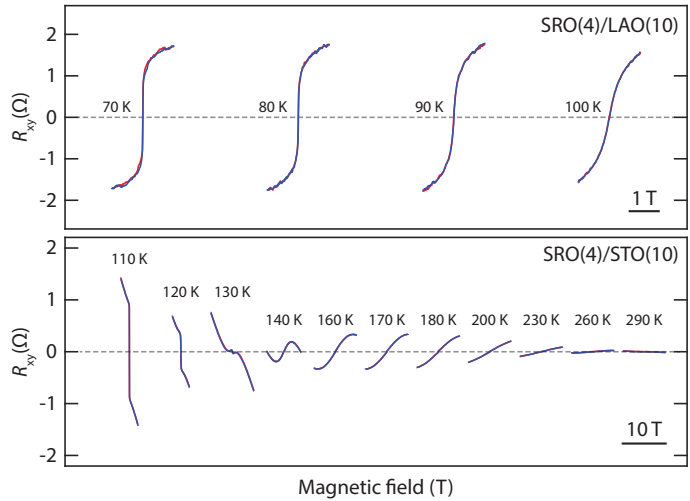

FIG S7. Hall effect as a function of magnetic field in the vicinity of the Curie temperature for (top) the SRO(4)/LAO and (bottom) the SRO(4)/STO heterostructure.

Therefore, the magneto-optical characterization likely slightly underestimates $T_C$ for all capping layers. Nevertheless, the suppression in $T_C$ between polar and non-polar samples is clearly observed by both techniques.

## II. DFT ANALYSIS OF THE LAO-SRO-STO HETEROSTRUCTURE

In this section we report the study of the LAO-SRO-STO heterostructure by considering the electronic properties and the derivation of the tight-binding electronic parameters for the effective $d$-bands for the Ti and Ru atoms.

### A. Methodology and structural properties

We perform spin polarized first-principles density functional calculations within the Local Spin Density Approximation (LSDA) by using the plane wave VASP DFT package [21, 22] and the Perdew-Zunger parametrization [23] of the Ceperley-Alder data [24] for the exchange-correlation functional. The choice of LSDA exchange functional is the most appropriate for the examined system because the Generalized Gradient Approximation has been shown to perform worse than LSDA for $SrRuO_3$ since it overestimates the amplitude of the magnetization [25]. The interaction between the core and the valence electrons was treated with the projector augmented wave (PAW) method [26] and a cutoff of 500 eV was used for the plane wave basis. The computational unit cells are constructed as supercells with three $SrTiO_3$ octahedra

having at least 20 Å of vacuum. Convergence was checked with respect to the thickness of the vacuum layer. The number of 6 layers is necessary to get the correct electric polarization behavior.

For Brillouin zone integrations, a $12 \times 12 \times 1$ k-point grid is used for the geometric relaxation and a $24 \times 24 \times 1$ k-point grid for the determination of the density of states (DOS), electron charge and magnetization distributions. A large number of k-points is necessary for an accurate determination of the magnetic moments in metallic multilayers. The atomic charge was calculated integrating in a sphere of 1.402 Å radius around the Ru atoms. The quantitative results are arbitrary because it depends on the size of the radius [27], the charge profile however is meaningful. We optimize the internal degrees of freedom by minimizing the total energy to be less than $3 \times 10^{-5}$ eV and the Hellmann-Feynman forces to be less than 20 meV/Å. The Coulomb interaction U at the Ru and Ti sites has been included in the LSDA+U approach using the rotational invariant scheme proposed by Liechtenstein [28]. We have used U=4 eV for the Ti $3d$ states while U=0.20 eV and $J_H$=0.15 U are considered for the Ru $4d$ states. In the literature, a value of the Coulomb repulsion U=0.50 eV was suggested for ruthenates to reproduce the magnetization of the bulk [29]. However, for the present analysis we employ a slightly smaller value because a lower amplitude of the magnetization is found for the metallic thin films [11].

The hopping parameters have been determined in the non magnetic phase. After obtaining the Bloch wave functions within the density functional theory, the maximally localized Wannier functions [30, 31] (MLWF) are constructed by means of the WANNIER90 computational scheme [32]. To extract the character of the electronic bands at low energies close to the Fermi level, we used the Slater-Koster interpolation scheme based on Wannier functions. The SRO thin films can be orthorhombic or tetragonal depending on the substrate, thickness, superlattice engineering and oxygen vacancies [33]. Here, we study the properties of different trilayers in the tetragonal phase without octahedral distortions with the $RuO_2$/LaO interface. We determine and compare the magnetic moment and the charge of the Ru atoms along the heterostructure for two sets of asymmetric trilayers $(STO)_3/(SRO)_n/(STO)_6$ and $(STO)_3/(SRO)_n/(LAO)_6$ with $n$=4,5,6. The results for $n$=4 are reported in Fig. 3 of the main text, the cases for n=5,6 are qualitatively equivalent with a Yukawa-like potantial when we move from the interface. The lack of octahedral rotations in the tetragonal phase tends to increase the strength of the $d-d$ hopping parameters with respect to the fully distorted SRO bulk structure. This results into a larger electronic bandwidth thus effectively reducing the electronic correlations and the magnetic moments along the heterostructure [34]. Therefore, the magnetic moment of the tetragonal phase turns out to be reduced if compared to the orthorhombic phase [35].

### B. Electronic structure and layer resolved density of states

We study the electronic properties of the LAO/SRO/STO interface in the tetragonal phase without octahedral distortions. The system is composed by a slab with 1 layer of STO, 1 layer of SRO and 2 layers of LAO. Two different interfaces $AlO_2$/SrO and LaO/$RuO_2$ are possible. Using the Wannier function method, we determine the effective hopping parameters for the $t_{2g}$ subsector of the Ti and Ru transition metal atoms. The Al states are far from the Fermi level due to the large gap of the LAO. [36]

We performed the structural relaxation of the slab for the two different interfaces. The in plane lattice constant is fixed to the value of the lattice constant of the STO. Our result for the value of the $c$-axis lattice constant is in qualitative agreement with the experimental value of 3.72 Å. The two different interfaces have different electronic and structural properties [36].

We have calculated the interface formation energy for the two possible interfaces $AlO_2/SrO$ and $LaO/RuO_2$. From the total energy calculation, we got that the interface formation energy is 2.8 eV for $AlO_2/SrO$ and 3.1 eV $LaO/RuO_2$. Therefore, the DFT suggests that the $LaO/RuO_2$ interface is more favorable to form because it optimizes the energy gain. The difficuly to experimentally achieve the $LaO/RuO_2$ surface might be related to the stability of the SrO termination of the $SrRuO_3$ [2, 37].

The band structure of the non magnetic systems is represented in Fig. S8. The band structure agrees with the reported band structure of tetragonal ruthenates and the hopping parameters are quantitatively similar to those previously reported for the ruthenates [16, 34]. Finally, we compute the layer resolved density of states for the Ru bands considering the magnetic phase of the $(STO)_3/(SRO)_4/(LAO)_6$ heterostructure. The results are reported in Fig. S9.

While the DFT calculations correctly reproduce qualitatively the experimental result of a reduced magnetization at the LAO/SRO interface, a quantitative comparison is more difficult to make because of the following considerations. First, we expect that the amount of charge reconstruction due to the polar interface can be underestimated in the LDA calculation [38]. This implies that the number of extra electrons filling the Ru bands is lower than the one experimentally achieved. Thus, the impact on the magnetization will be less pronounced.

Second, the description of the ferromagnetic ground state is primarily based on the Stoner mechanism (see also [19]) and it does not take into account spin fluctuations. Due to the suppression of the magnetic moment at the interface layer, one expects that the spin fluctuations will be stronger and more relevant in setting out the amplitude of the magnetization. Such spin modes are also effective to drive a renormalization of the magnetic moment in the inner layers.

Third, due to the valence mismatch La intermixing (i.e., sporadic A-site substitution of Sr by La) produces spatial charge (and spin) fluctuations. The LDA analysis assumes translational invariance and ideal matching at the interface. Therefore, in the presence of intermixing, this method is expected to underestimate the reduction of the magnetization..

### C. Tight-binding model by the maximally localized Wannier approach

A tight-binding model is derived including the nearest-neighbor (NN) and next-nearest-neighbor (NNN) hopping terms without the atomic spin-orbit coupling that will be included afterwards for the Hall conductivity computation. We restrict our analysis to the effective $d$-bands of the Ti and Ru atoms. The simulated system is made with 2 metal atoms in the unit cell, and we write the tight-binding model for a bilayer without performing the Fourier transform along the out-of-plane $z$-direction. We label the two atoms of the bilayer as 1 and 2. The Hamiltonian in matrix form for the bilayer is expressed as:

$$\hat{H}(k_x, k_y) = \begin{pmatrix} H_{11}(k_x, k_y) & H_{12}(k_x, k_y) \\ H_{21}(k_x, k_y) & H_{22}(k_x, k_y) \end{pmatrix}$$

Each submatrix is a sum of the different electronic processes related to the energy on site, NN and NNN hoppings as $H_{11} = H_{11}^0 + H_{11}^{NN} + H_{11}^{NNN}$

$$H_{11}^0 = \begin{pmatrix} \varepsilon_{xy}^0 & 0 & 0 \\ 0 & \varepsilon_{xz}^0 & 0 \\ 0 & 0 & \varepsilon_{yz}^0 \end{pmatrix}$$

$$H_{11}^{NN} = \begin{pmatrix} 2t_{xy,xy}^{100}\cos k_x a + 2t_{xy,xy}^{010}\cos k_y a & 0 & 0 \\ 0 & 2t_{xz,xz}^{100}\cos k_x a + 2t_{xz,xz}^{010}\cos k_y a & 0 \\ 0 & 0 & 2t_{yz,yz}^{100}\cos k_x a + 2t_{yz,yz}^{010}\cos k_y a \end{pmatrix}$$

$$H_{11}^{NNN} = \begin{pmatrix} 4t_{xy,xy}^{110}\cos k_x a \cos k_y a & 0 & 0 \\ 0 & 4t_{xz,xz}^{110}\cos k_x a \cos k_y a & -4t_{xz,yz}^{110}\sin k_x a \sin k_y a \\ 0 & -4t_{xz,yz}^{110}\sin k_x a \sin k_y a & 4t_{yz,yz}^{110}\cos k_x a \cos k_y a \end{pmatrix}$$

while for the submatrix $H_{12}$ connecting the two layers we have:

$$H_{12}^{NN} = \begin{pmatrix} t_{xy,xy}^{001} & 0 & 0 \\ 0 & t_{xz,xz}^{001} & 0 \\ 0 & 0 & t_{yz,yz}^{001} \end{pmatrix}$$

$$H_{12}^{NNN} = \begin{pmatrix} 2t_{xy,xy}^{101}\cos k_x a + 2t_{xy,xy}^{011}\cos k_y a & 2it_{xy,xz}^{011}\sin k_y a & 2it_{xy,yz}^{101}\sin k_x a \\ 2it_{xz,xy}^{011}\sin k_y a & 2t_{xz,xz}^{101}\cos k_x a + 2t_{xz,xz}^{011}\cos k_y a & 0 \\ 2it_{yz,xy}^{101}\sin k_x a & 0 & 2t_{yz,yz}^{101}\cos k_x a + 2t_{yz,yz}^{011}\cos k_y a \end{pmatrix}$$

The breaking of the inversion symmetry along $z$ (BISz) gives rise to spin-independent hopping terms that mix $xy$ and $\gamma z$ orbitals also for the NN. This type of hopping is responsible for the so-called orbital Rashba effect [39]. In the case of the STO/SRO and SIO/SRO interfaces such terms are negligible, being of the order of few meV and modifying the band structure half eV above the Fermi level. However, in the case of the LAO/SRO/STO these orbital Rashba terms are of the order of 10 meV due to the large breaking of the inversion symmetry produced by the LAO layer. The Al atoms are far from the Fermi level, therefore we consider only the Ru and Ti $d$-states in our Wannier function basis set.

$$H_{11}^{NN,BISz} = \begin{pmatrix} 0 & 2it_{xy,xz}^{010}\sin k_y a & 2it_{xy,yz}^{100}\sin k_x a \\ -2it_{xy,xz}^{010}\sin k_y a & 0 & 0 \\ -2it_{xy,yz}^{100}\sin k_x a & 0 & 0 \end{pmatrix}$$

$$H_{11}^{NNN,BISz} = \begin{pmatrix} 0 & 4it_{xy,xz}^{110}\sin k_y a \cos k_x a & 4it_{xy,yz}^{110}\sin k_x a \cos k_y a \\ -4it_{xy,xz}^{110}\sin k_y a \cos k_x a & 0 & 0 \\ -4it_{xy,yz}^{110}\sin k_x a \cos k_y a & 0 & 0 \end{pmatrix}$$

We list the hopping parameters between Wannier functions (WF) of the same species in Table I and between inequivalent species in Table II. In the tetragonal case there is no first neighbour hopping between different orbitals. Non diagonal hoppings can be activated by the octahedral rotations or breaking of the inversion symmetry as in this case. The two main differences between the $t_{2g}$ Ti states and the $t_{2g}$ Ru states are the position of the energetic levels and the size of the hopping. Due to the reduced atomic radius of the Ti the hopping of the Ti orbitals are smaller and the energetic level of the Ti orbital are about 2 eV higher than the Ru energetic levels. The Ru and Ti bands are well separated in energy as shown in Figure S8, while in the SrRuO$_3$/SrIrO$_3$ superlattice there is a strong mixing between the $d$-states of the different atoms. We list the hopping parameters between the same atomic species in Table III and the mixing terms in Table IV. We observe that the xz(Ti)-yz(Ti) hopping along 110 is much smaller than the xz(Ru)-yz(Ru) amplitude. We can separate the hopping parameters in 4 groups. Just above the Fermi level, the $e_g$ of the Ru are also visible while the La states are around 3-4 eV above the Fermi level depending on the kind of interface. The n-type interface is the LaO/TiO$_2$ while the p-type is AlO$_2$/SrO. The LaO/RuO$_2$ interface presents larger Rashba terms for the Ru atoms since the Ru atoms are closer to the LaAlO$_3$.

The strength of the orbital Rashba depends on the quantity $\frac{t_{xy,yz}^{100}}{t_{xz,xz}^{100}}$. This ratio is 0.030 for the RuO$_2$/LaO interface while it is 0.004 for the SrO/AlO$_2$ interface. The RuO$_2$/LaO shows an orbital Rashba that is one order of magnitude larger than the other interface. To evaluate the effects of the orbital Rashba in this system, taking also into account the experimental inputs, we focus on the RuO$_2$/LaO interface. The numerical values of the orbital Rashba used in the model Hamiltonian are of the order of magnitude of the values obtained in DFT.

### D. Density of states at the planar and apical oxygen sites

We determine the density of states (DOS) projected on the planar and apical oxygens for the in-plane ($p_{x/y}$) and out-of-plane ($p_z$) orbital polarizations for both the LAO/SRO/STO and STO/SRO/STO heterostructures. This analysis can be directly relevant for quantitatively assessing the occurrence of charge reconstruction at the LAO/SRO interface by the use of O $1s$ x-ray-absorption spectroscopy (XAS) or electron energy loss spectroscopy (EELS). The latter technique could provide a spatial profile of the charge state across the heterostructure. However, direct measurement of the Ru $L$-edge is typically inaccessible due to its high energy $\sim 3$ keV. Instead, one should look at the O $K$-edge, which has energy $\sim 0.5$ keV. The electric field of incoming photons for the spectroscopic detection, due to the dipole selection rules, leads to O $1s \rightarrow$ O $2\, p_{x/y}$ excitations for an in-plane orientation, while the out-of-plane

TABLE I. Hopping integrals between the M atoms for the nearest-neighbor (NN), for the next-nearest-neighbor (NNN) as the selected WFs of LaO/RuO$_2$ between WFs of the same atomic specie. The on-site energy of the Ru xy-like WF is set to zero. The unit is meV.

|  | on site | NN | | NNN |
|---|---|---|---|---|
|  | 000 | 100 | 010 | 110 |
| xy(Ti)-xy(Ti) | 2242 | -286 | -286 | -86 |
| xz(Ti)-xz(Ti) | 2430 | -218 | -52 | 6 |
| yz(Ti)-yz(Ti) | 2430 | -52 | -218 | 6 |
| xz(Ti)-yz(Ti) | 0 | 0 | 0 | 4 |
| xy(Ti)-xz(Ti)=$\lambda_R$ | 0 | 0 | 13 | -4 |
| xy(Ti)-yz(Ti)=$\lambda_R$ | 0 | 13 | 0 | -4 |
| xy(Ru)-xy(Ru) | 0 | -352 | -352 | -124 |
| xz(Ru)-xz(Ru) | 235 | -265 | -17 | 17 |
| yz(Ru)-yz(Ru) | 235 | -17 | -265 | 17 |
| xz(Ru)-yz(Ru) | 0 | 0 | 0 | 18 |
| xy(Ru)-xz(Ru)=$\lambda_R$ | 0 | 0 | 8 | -3 |
| xy(Ru)-yz(Ru)=$\lambda_R$ | 0 | 8 | 0 | -3 |

TABLE II. Hopping integrals between the M atoms for the nearest-neighbor (NN), for the next-nearest-neighbor (NNN) as the selected WFs of LaO/RuO$_2$ between WFs of different atomic species. The unit is meV.

|  | NN | NNN | |
|---|---|---|---|
|  | 001 | 101 | 011 |
| xy(Ru)-xy(Ti) | -36 | 12 | 12 |
| xz(Ru)-xz(Ti) | -318 | -88 | 10 |
| yz(Ru)-yz(Ti) | -318 | 10 | -88 |
| xy(Ru)-xz(Ti) | 0 | 0 | 12 |
| xy(Ti)-xz(Ru) | 0 | 0 | 9 |

configuration induces exclusively O $1s \rightarrow$ O $2\,p_z$ transitions. Moreover, the intensities of the low-energy features in the absorption spectra within a window of about 1 eV from the Fermi energy reflect the strength of the Ru $4d$-O $2p$ orbital hybridization for both in-plane and apical oxygens [40, 41]. Thus, we expect that in this range of energy excitations the charge reconstruction for the Ru $4d$ states at the LAO/SRO interface (Fig. 1) would have consequences on the low-energy unoccupied DOS of the O $2p$

TABLE III. Hopping integrals between the transition metal atoms for the nearest-neighbor (NN), for the next-nearest-neighbor (NNN) as the selected WFs of AlO$_2$/SrO between WFs of the same atomic species. The on-site energy of the Ru xy-like WF is set to zero. The unit is meV.

|  | on site | NN | | NNN |
|---|---|---|---|---|
|  | 000 | 100 | 010 | 110 |
| xy(Ti)-xy(Ti) | 2257 | -287 | -287 | -87 |
| xz(Ti)-xz(Ti) | 2438 | -214 | -55 | 4 |
| yz(Ti)-yz(Ti) | 2438 | -55 | -214 | 4 |
| xz(Ti)-yz(Ti) | 0 | 0 | 0 | 3 |
| xy(Ti)-xz(Ti)=$\lambda_R$ | 0 | 0 | 6 | -4 |
| xy(Ti)-yz(Ti)=$\lambda_R$ | 0 | 6 | 0 | -4 |
| xy(Ru)-xy(Ru) | 0 | -354 | -354 | -123 |
| xz(Ru)-xz(Ru) | 221 | -277 | -32 | 14 |
| yz(Ru)-yz(Ru) | 221 | -32 | -277 | 14 |
| xz(Ru)-yz(Ru) | 0 | 0 | 0 | 16 |
| xy(Ru)-xz(Ru)=$\lambda_R$ | 0 | 0 | 1 | 0.5 |
| xy(Ru)-yz(Ru)=$\lambda_R$ | 0 | 1 | 0 | 0.5 |

TABLE IV. Hopping integrals between the M atoms for the nearest-neighbor (NN), for the next-nearest-neighbor (NNN) as the selected WFs of AlO$_2$/SrO between WFs of different atomic species. The unit is meV.

|  | NN | NNN | |
|---|---|---|---|
|  | 001 | 101 | 011 |
| xy(Ru)-xy(Ti) | -36 | 12 | 12 |
| xz(Ru)-xz(Ti) | -315 | -91 | 5 |
| yz(Ru)-yz(Ti) | -315 | 5 | -91 |
| xy(Ru)-xz(Ti) | 0 | 0 | 14 |
| xy(Ti)-xz(Ru) | 0 | 0 | 7 |

bands.

To this aim, we focus on the DOS for the empty states above the Fermi level considering a representative LAO/SRO/STO heterostructure (Fig. S10) with six SRO unit cells. Such a heterostructure has been also considered for the investigation of the transport properties. Moreover, to evaluate the impact of the interface reconstruction due to the LAO, we directly compare the DOS for the STO/SRO/STO

and LAO/SRO/STO heterostructures. Since the structural reconstruction is different for STO/SRO and LAO/SRO interfaces, in order to disentangle the effect of the charge reconstruction from the buckling, we

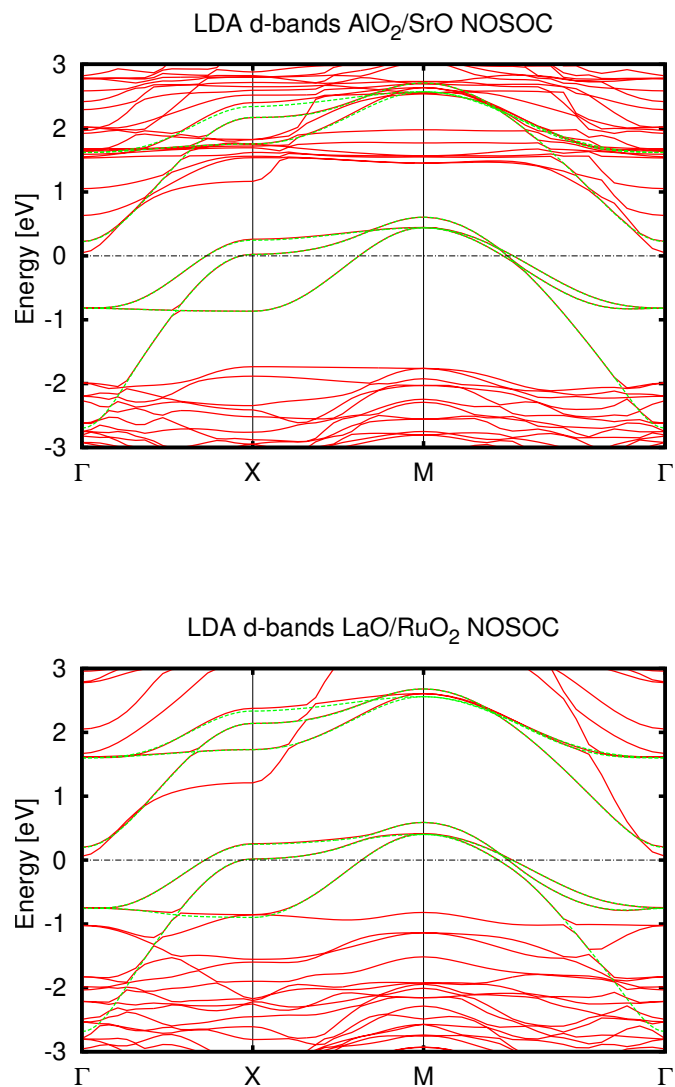

FIG S8. DFT band structure (red) and $t_{2g}$ bands (green) obtained using the Wannier functions for the nonmagnetic phase of $AlO_2$/SrO (left panel) and LaO/$RuO_2$ (right panel) superlattice. The Fermi level is set to zero.

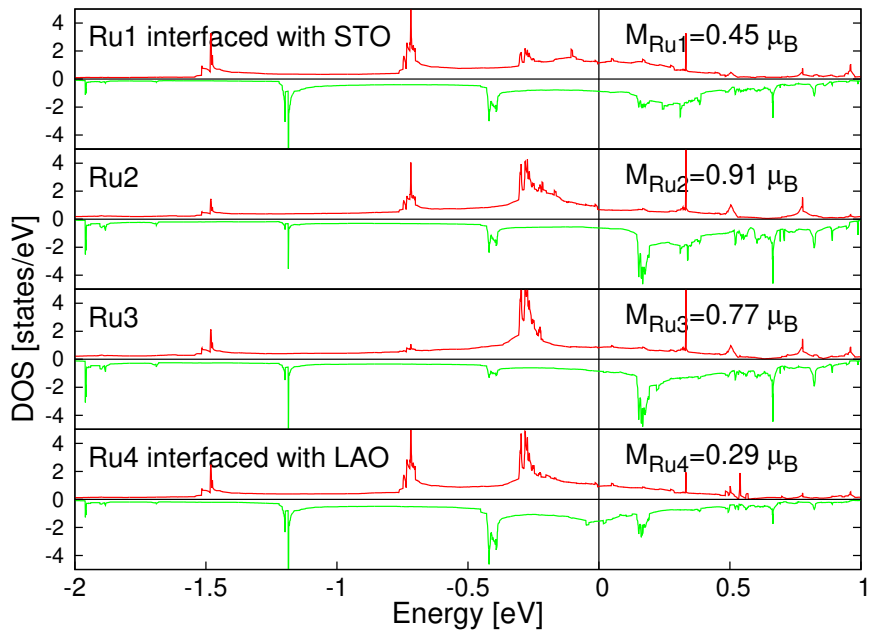

FIG S9. Layer resolved DOS for the Ru atoms of the $(STO)_3/(SRO)_4/(LAO)_6$ heterostructures for U=0.2 eV. We indicate the local magnetization of the Ru atoms. Majority (minority) spin is represented in the upper (lower) half of the panels.

consider the STO/SRO/STO atomic positions for both LAO/SRO/STO and STO/SRO/STO heterostructures. Indeed, the computation has been also performed for the STO/SRO/STO heterostructure to have a reference where there is no charge reconstruction at the interface.

The evolution of the DOS for the planar and apical oxygens is presented in Figs. S11 and S12, respectively. We find that the largest part of the spectral weight of the unoccupied states is between zero and about 0.4 eV where the Ru-$t_{2g}$ orbitals are placed. In this energy range, we observe that the major difference between the STO/SRO/STO and LAO/SRO/STO heterostructure occurs for the first two unit cells (i.e. for u.c.i=5,6) close to the LAO interface. For those unit cells, indeed, we expect that the structural and charge reconstructions are mostly affecting the DOS. When comparing the DOS for the STO/SRO/STO and LAO/SRO/STO heterostructures, the results indicate a reduction and slight shift of the spectral weight from about 0.3 eV to 0.1 eV for the planar DOS (Fig. S11) while for the apical DOS only a redistribution of intensity is observed (Fig. S12). For the remaining unit cells the differences between the STO/SRO/STO and LAO/SRO/STO heterostructures is negligible. Concerning the orbital polarization dependence of the DOS, we do not find significant variations for the $p_{x/y}$ and $p_z$

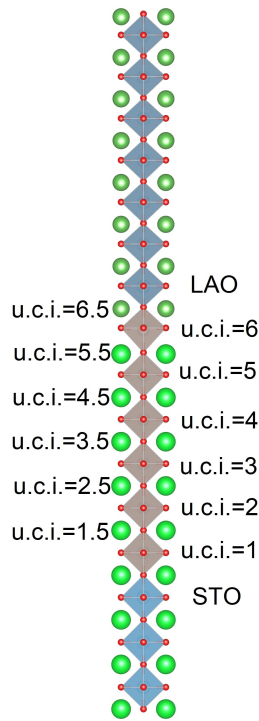

FIG S10. Sketch of the LAO/SRO/STO heterostructure with six SRO unit cell that has been employed for the computation of the DOS. The unit cell index (u.c.i) indicates the RuO$_2$ layers for integer values, while for half-integer amplitudes the SrO layers and the LaO interface layer, respectively.

projected DOS concerning the planar oxygens. On the other hand, due to the structure of the Ru 4$d$-O 2$p$ hybridization, the intensity of the spectral weight for the apical oxygens is nonvanishing only for the $p_{x/y}$ orbitals.

The theoretical analysis provides clear indications about the impact of structural and charge reconstruction at the LAO/SRO interface compared with the STO/SRO electronic configuration. The main variations for the low-energy excitations in the DOS manifest at the single unit-cell level, in a range of energy of the order of $0.2\,\mathrm{eV}$ with rather broad features. While this is beyond the capabilities of a typical STEM-EELS setup to detect, more advanced techniques might be able to resolve such features.

## III. BERRY CURVATURE AND ANOMALOUS HALL CONDUCTIVITY IN LAO-SRO-STO

In this section we analyze the modification of the topological properties of the Ru $d$-bands for the LAO-SRO-STO heterostructure. In doing that we take into account the inputs from the study and the

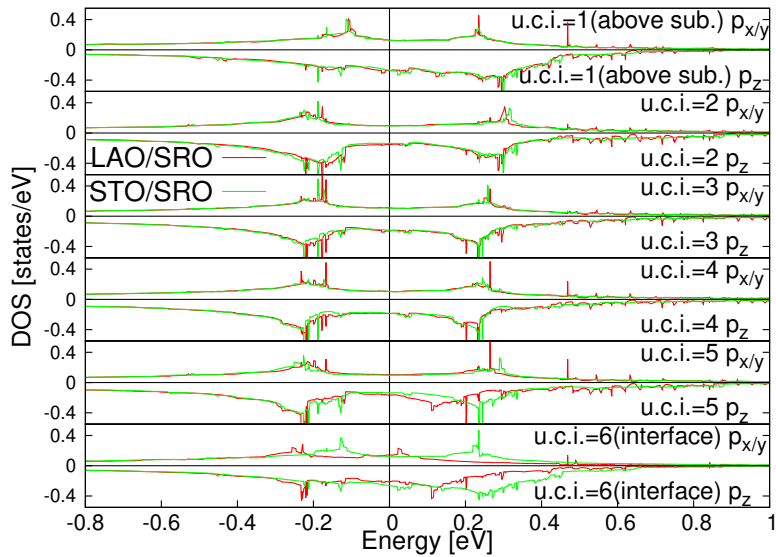

FIG S11. Evolution of the DOS for the planar oxygens with $p_{x/y}$ and $p_z$ configurations for the LAO/SRO/STO (red) and STO/SRO/STO (green) heterostructures, respectively.

derived tight-binding model. In particular, we simulate the presence of the LAO polar interface as a source of inversion symmetry breaking and charge reconstruction of the Ru bands occupation at the SRO interface layer. Then, for the resulting electronic structure we determine the topological contribution to the Hall conductivity due to the Berry curvature of the Ru bands in the momentum space for ferromagnetic SrRuO$_3$ ultra-thin films for different types of electronically reconstructed interface. Here, we recall that the Ru bands in the two-dimensional monolayer limit have a topological character with nontrival Chern number that can be associated to the $t_{2g}$ spin-orbital mixed bands [16].

The aim here is to evaluate the impact on the anomalous Hall conductance (AHC) due to the charge reconstruction and electric field effects that arise in the Ru $t_{2g}$ bands of SRO in the presence of LAO polar interface. The study is performed by varying the magnetization of the ferromagnet to assess the role of the spin polarization strength. In particular, the calculation is done for a bilayered heterostructure with two Ru-O planes that we label with the index $i_z = 1$ and $i_z = 2$. This is a minimal configuration that allows us to directly assess the role of both electron doping of Ru bands and inversion symmetry breaking induced at polar interface. The reference case is to have both layers interfaced with insulating electronic systems (i.e. SrTiO$_3$) assuming a uniform charge density of four electrons per Ru atoms as expected for the Ru$^{4+}$ valence. All the electronic parameters are taken from the LDA and Wannier projection analysis as described in the previous section. In particular, for clarity the unit scale $t$ is the amplitude of the in-

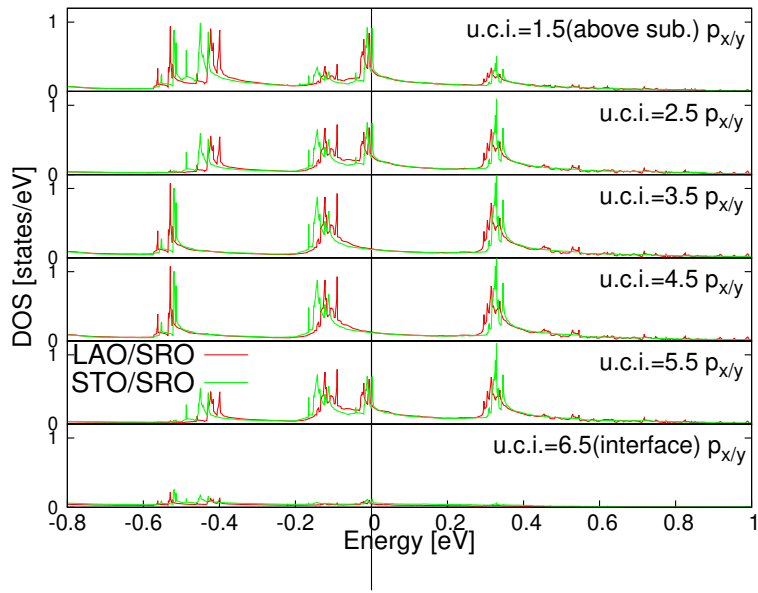

FIG S12. Evolution of the DOS for the apical oxygens with $p_{x/y}$ polarization for the LAO/SRO/STO (red) and STO/SRO/STO (green) heterostructures, respectively.

plane nearest neighbor $d - d$ hopping $t_1$, while in-plane second nearest-neighbor $t_2 = -0.16\,t$, atomic spin-orbit $\lambda_{so} = -0.2\,t$, and the interlayer hopping $t_\perp = 0.5\,t$.

For the reference configuration, the anomalous Hall conductivity is always negative for any value of the magnetization ranging from 0 to 1 $\mu_B$/Ru as shown in Fig. S13.

To proceed further, we present the evolution of the anomalous Hall conductance which is analyzed as a function of the magnetization assuming that the spin-polarization is oriented out-of-plane ($\hat{z}$-direction). Here, according to the analysis, we explore different electronic configurations that can be induced by charge-reconstruction interface effects. Indeed, assuming that the top layer ($i_z = 2$) of the bilayer is interfaced with a polar interface (i.e. LAO), we expect that, due to the La/Sr valence change and the electric field generated by the La-O layers, there occurs a charge redistribution together with the emergence of an inversion symmetry breaking interaction. In this context, the relevant electronic parameters of the model Hamiltonian are: i) the orbital Rashba coupling $\lambda^R$ (induced by the electric field due to the polar interface), ii) the charge potential $V$ to set the charge imbalance between the interface layer and the nearby Ru layer, and iii) the inter-layer hopping $t_\perp$ to control the degree of charge proximity between the two layers. We recall that the orbital Rashba coupling by symmetry is proportional to the interface electric field independently of structural distorsions and the corresponding term in the Hamiltonian is conventionally expressed as $H_{i_z}^R = \lambda_{i_z}^R \left[ \sin(k_y)\hat{L}_x - \sin(k_x)\hat{L}_y \right]$. $\hat{L}_i$ are the components of the effective

angular momentum in the $t_{2g}$ manifold.

Due to the La/Sr valence variation induced at the La-O interface layer, the electron density for Ru is expected to be 4.5 and thus we have that the total electron filling for unit cell is $N = 8.5$. Let us start by considering the AH modification due to the total electron filling in the presence of a non-vanishing orbital Rashba coupling at the La-O interface layer, i.e. $\lambda_1^R = 0$ and $\lambda_2^R \neq 0$, neglecting the charge potential and the renormalization of the inter-layer charge transfer. In this case, we find that the AH conductance is only slightly modified with respect to the reference case because the excess charge almost equally distributes in the two Ru layers. The main effect of the orbital Rashba coupling in this configuration is to shift the maximum of the AH conductance to a lower amplitude of the magnetization. This is plausible because the orbital Rashba coupling breaks inversion and thus induces other electronic channels of nontrivial Berry curvature associated with an orbital polarization oriented in the $xy$-plane that can interfere with the Berry curvature distribution due to the ferromagnetic spin-polarization.

Then, we allow for an increasing charge imbalance between the two Ru layers by modifying both the charge potential at the interface, $V_2$, and the amplitude of the inter-layer charge transfer through the factor $\gamma$ such as $t'_{\perp} = \gamma\, t_{\perp}$. The analysis is done by keeping fixed the total electron density in the unit cell which, for the case examined is $N = 8.5$. Both mechanisms lead to a localization of the excess charge of 0.5 electron per ruthenium atom within the Ru-O layer nearby the polar interface. As we demonstrate in Fig. S13, the overall trend of the AH conductance is quite clear. The AH conductance tends to change

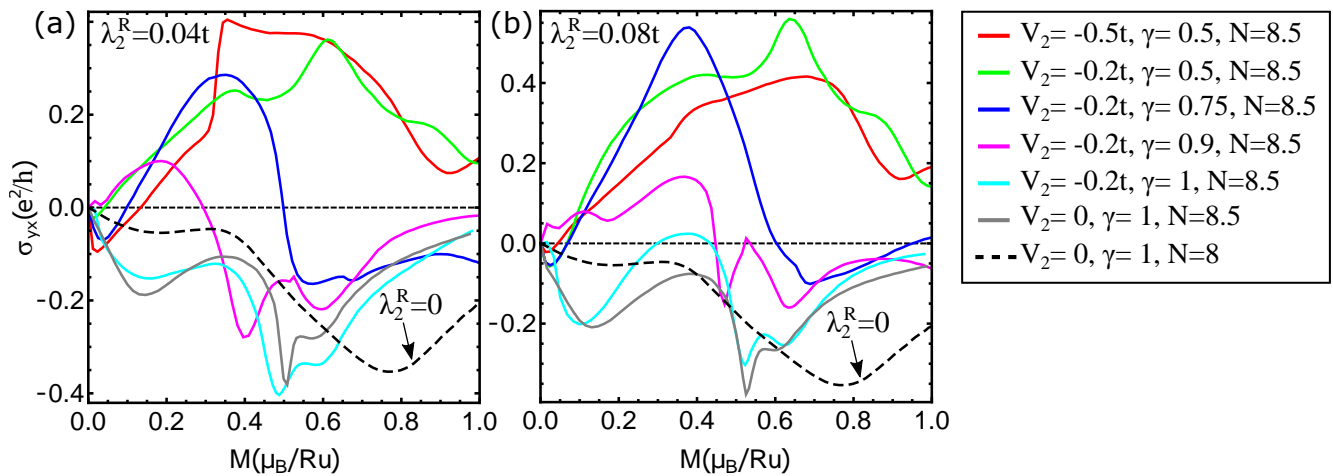

FIG S13. AH conductance curves versus magnetization $M$ for Ru-Ru bilayer and different electronic configurations. The unit cell is made of two ruthenium atoms and we assume a fixed electron density $N$. We consider two representative values of the orbital Rashba coupling at the interface, (a) $\lambda_2^R = 0.04\, t$ and (b) $\lambda_2^R = 0.08\, t$ and vary the total electron filling from $N = 8$ to $N = 8.5$. The reference case is without charge reconstruction and vanishing orbital Rashba interaction. The other parameters are: in-plane first nearest-neighbor $t_1 = -t$ and second nearest-neighbor $t_2 = -0.16t$, atomic SOC $\lambda = -0.2t$.

sign as one increases the localization of the excess charge at the top layer. The sign change firstly occurs in the regime of low magnetization as the AH conductance is vanishing and thus the fluctuations are more significant. The switching to a fully positive AH conductance can be driven both by a renormalization of the inter-layer charge transfer or by considering the charge potential that pins the electron density at the interface to be as much as possible close to $4.5$. An additional remark concerns the role of the orbital Rashba coupling. We find that a larger amplitude of the orbital Rashba coupling allows to extend the sign change of the AH conductance in almost the whole range of examined magnetizations. For instance, considering the case with $V_2 = -0.2\,t$ and $\gamma = 0.75$, the case with $\lambda_2^R = 0.08\,t$ (Fig. S13(b)) the AH conductance is positive for all values of the magnetization except in the range $\sim[0.65\text{-}1.0]\,\mu_B$. Instead, for $\lambda_2^R = 0.04\,t$, with the same values of $V_2$ and $\gamma$ the sign change is obtained only for values of $M_{Ru}$ in the range $[0.0\text{-}0.5]\,\mu_B$.

The overall analysis indicates that the charge reconstruction and the electric field at the polar interface generally induces a sign change in the AH conductance which is, however, dependent of the strength of the magnetization and on the degree of localization of the interface charge.

It is interesting to evaluate whether the sign reconstruction of the anomalous Hall conductance can be understood as a direct sum of contributions arising from the layer projected Hall conductance. Since the charge reconstruction substantially modifies the electronic configuration only for the Ru bands at the polar interface layer, one may argue that those states yield a conductance with an opposite sign with respect to the electronic bands on the other interface. For doing that one can use the Kubo formula and decompose the velocity operator to assess the contributions from each layer. Indeed, the total intrinsic AH conductance can be expressed as

$$\sigma_{tot} = \frac{e^2}{h}\frac{1}{2\pi}\sum_{n\neq n'}\int d^2k\,(f_{k,n} - f_{k,n'})\,\mathrm{Im}\frac{\langle k,n|\,\partial_{k_x}\mathcal{H}\,|k,n'\rangle\,\langle k,n'|\,\partial_{k_y}\mathcal{H}\,|k,n\rangle}{\left(E_{\vec{k}}^{(n)} - E_{\vec{k}}^{(n')}\right)^2}.$$

The layer-projected AHE conductances $\sigma_1$ and $\sigma_2$ are given by

$$\sigma_i = \frac{e^2}{h}\frac{1}{2\pi}\sum_{n\neq n'}\int d^2k\,(f_{k,n} - f_{k,n'})\,\mathrm{Im}\frac{\langle k,n|\,\partial_{k_x}\mathcal{H}_i\,|k,n'\rangle\,\langle k,n'|\,\partial_{k_y}\mathcal{H}_i\,|k,n\rangle}{\left(E_{\vec{k}}^{(n)} - E_{\vec{k}}^{(n')}\right)^2}$$

where

$$\mathcal{H}_i = P_i\mathcal{H}P_i$$

and $P_i$ is a projector on the layer $i_z = 1, 2$, respectively. Then, by combining the amplitude of the total conductance with that of the individual layers, one can deduce a projector $\sigma_{12} = \sigma_{tot} - \sigma_1 - \sigma_2$ that takes into account of the inter-layer interference. The results are reported in Fig. S14. As one can judge from the overall outcome, the total AHC can be hardly decomposed into the contributions arising only from the individual layers. The interference or inter-layer term is typically crucial to set the total AHC. This behavior is not completely unexpected if one observes that the electron density is not the only parameter entering into the electric current response. Then, even if one is producing a strong charge unbalance

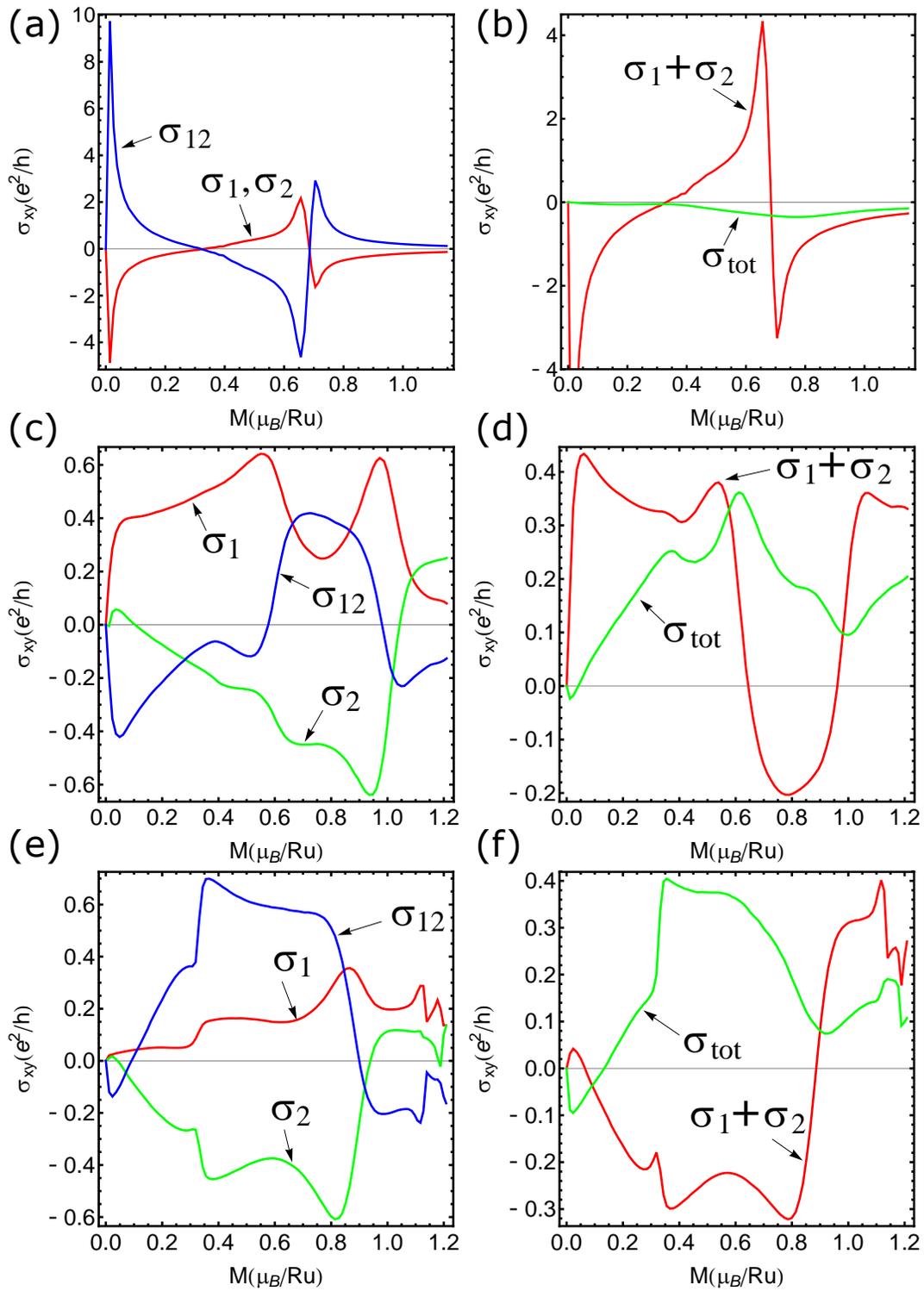

FIG S14. AHE curves versus magnetization $M$ for a Ru-Ru bilayer decomposed into different channels; $\sigma_1$ and $\sigma_2$ are the conductance projected on the first and second layer, $\sigma_{tot}$ is the total AH conductance and $\sigma_{12}$ is the derived inter-layer contribution as given by $\sigma_{12} = \sigma_{tot} - \sigma_1 - \sigma_2$. The parameters are: (a-b) $V_2 = 0$, $\gamma = 1$, $N = 8$ (symmetric Ru-Ru case), (c-d) $V_2 = -0.2t$, $\gamma = 0.5$, $N = 8.5$ and (e-f) $V_2 = -0.5t$, $\gamma = 0.5$, $N = 8.5$. The other electronic parameters are: in-plane first and second NNs, $t_1 = -t$, $t_2 = -0.16t$, atomic SOC $\lambda = -0.2t$.

between the layers, the total AHC keeps showing a subtle dependence on the inter-layer terms which is a sort of manifestation of the current-current correlations between electrons flowing in different layers. In this respect, one can argue that even in the presence of a complete charge reconstruction uniquely at the interface layer, the inter-layer term is not negligible and this quantum interference is quite relevant. Indeed, in the panel (f) of Fig. S14 the sum of the AHC for each separate layer has an opposite sign with respect to the total one. This behavior is obtained almost in the whole range of relevant magnetization amplitudes. The sum two decoupled Ru layers with different electron filling cannot account for the sign change. This theoretical result can have implications on the Berry phase because the electronic coupling between the Ru layers is relevant to yield a sign reconstruction of the AHC.

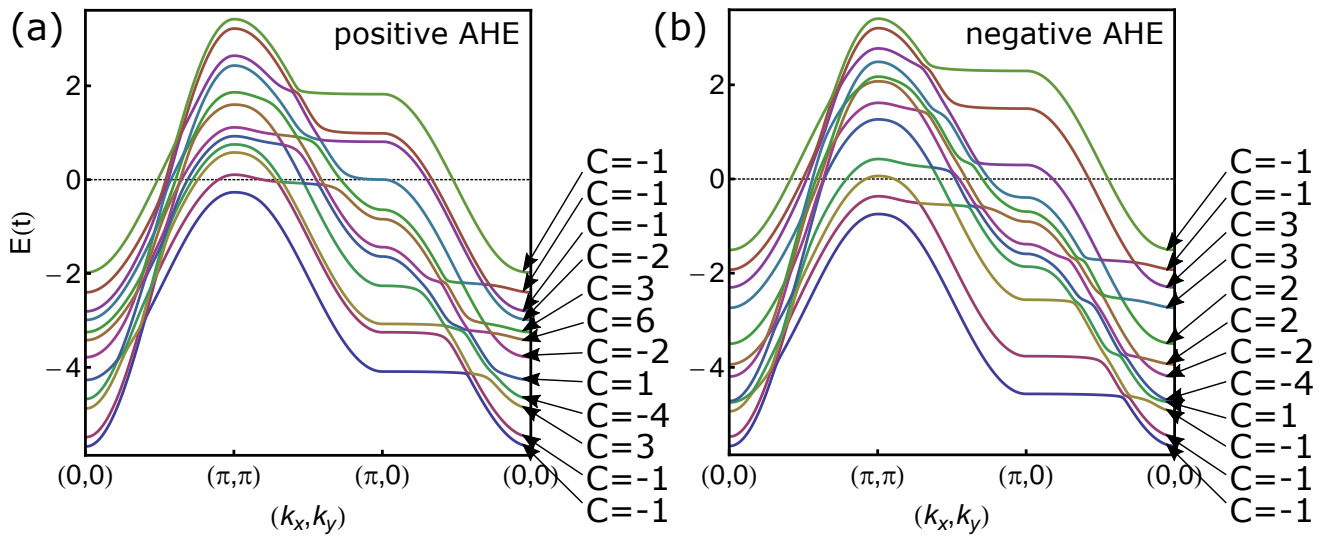

FIG S15. Band structures and band-resolved Chern numbers for the case of (a) strong charge pinning and (b) weak charge pinning, at magnetization $M = 0.5\ \mu_B/$Ru.

We proceed with a derivation that can be employed to extract information on the behavior of the anomalous Hall conductance, taking mainly into account the emerging topological character of the electronic bands expressed by the value of their Chern number. The conductance $\sigma_{xy}$ can generally be expressed as an integration of the Berry curvature, weighted by the occupation number of the corresponding band i.e.,

$$\sigma_{xy} = \frac{e^2}{h} \sum_{n=1}^{d} \int_{BZ} d^2k f\left(E_{\vec{k}}^{(n)}\right) \Omega_{\vec{k}}^{(n)} \qquad (2)$$

where $f$ is the Fermi-Dirac distribution, and the Berry curvature $\Omega_{\vec{k}}^{(n)}$ that is defined as

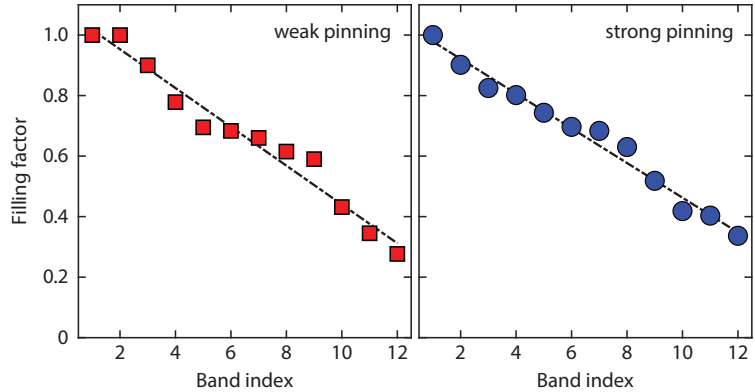

FIG S16. Filling factors determined from the electronic structure for (left) the weak and (right) the strong charge pinning scenarios. The band index represents the bands sorted in ascending order of energy. The dashed lines represent linear fits.

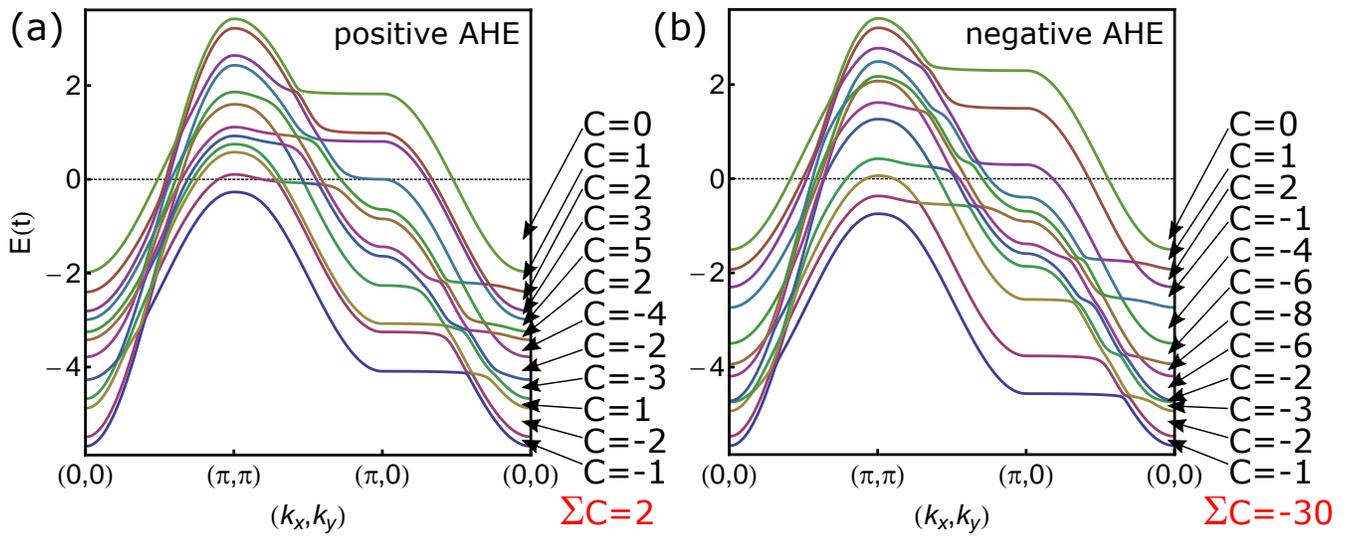

FIG S17. Band structures and Chern numbers associated with the indirect gaps for the case of (a) strong charge pinning and (b) weak charge pinning, at magnetization $M = 0.5 \, \mu_B/$Ru. The sum of the Chern numbers is denoted in red, transitioning from positive to negative between the two scenarios.

$$\Omega_{\vec{k}}^{(n)} = \frac{1}{\pi} \sum_{\substack{n'=1 \\ n' \neq n}}^{d} Im \frac{\left\langle n, \vec{k} \right| \partial_{k_x} \mathcal{H} \left| n', \vec{k} \right\rangle \left\langle n', \vec{k} \right| \partial_{k_y} \mathcal{H} \left| n, \vec{k} \right\rangle}{\left( E_{\vec{k}}^{(n)} - E_{\vec{k}}^{(n')} \right)^2}. \tag{3}$$

One can determine the Chern number for each given band $n$ by integrating the Berry curvature over the full Brillouin zone (see Fig. S15), i.e.,

$$\mathcal{C}_n = \int_{BZ} d^2 k \, \Omega_{\vec{k}}^{(n)}. \tag{4}$$

Now one can obtain an insightful estimation of the anomalous Hall conductivity, by considering the momentum-space averaged Berry curvature, which is directly proportional to its Chern number as

$$\Omega_{\vec{k}}^{(n)} \simeq \frac{1}{(2\pi)^2} \mathcal{C}^{(n)}. \tag{5}$$

We note that this zeroth-order approximation is expected to fail for a single band, because the Berry curvature typically has sharp peaks at the minima of the indirect gaps and is otherwise small (i.e., higher-order terms may be non-negligible). In an averaged computation over all bands however, the approximation becomes more accurate as higher-order terms can interfere destructively, thereby making the zeroth-order term more dominant. Additionally, it is expected that external mechanisms such as scattering and disorder will further cause the Berry curvature to smoothen in momentum space. Then, one can proceed by approximating the conductivity $\sigma_{xy}$ as

$$\sigma_{xy} \simeq \frac{e^2}{h} \frac{1}{(2\pi)^2} \sum_{n=1}^{d} A_n \mathcal{C}_n, \tag{6}$$

where $A_n = \int_{BZ} d^2 k \, f\left(E_{\vec{k}}^{(n)}\right)$ is the occupied fraction of $n$. While the exact filling factors have to be explicitly computed, it is anticipated that if the bands are ordered in such a way that $n = 1$ is the lowest one in energy, $n = 2$ the second, etc., then the filling factor $A_n$ decreases with $n$. The simplest form for $A_n$ would be to assume the first band as the most occupied and the others evolving with a linearly decreasing profile. Comparing linear fits with numerically computed filling factors (Fig. S16), we find that this approximation is justified. Thus, one can express $A_n$ as

$$A_n \simeq (2\pi)^2 (\alpha - \beta n) \tag{7}$$

with $\alpha$ and $\beta$ generic positive coefficients that model the linear dependence of the band occupation, with the constraint that $\alpha > \beta d$. Hence, let us introduce the quantity $F_n$ that is expressed as

$$F_n = \sum_{n'=1}^{n} \mathcal{C}_{n'} \tag{8}$$

and gives the sum of the Chern numbers up to the band $n$, or equivalently the Chern number associated with the indirect gap between band $n$ and $n + 1$. With the above expressions, the equation 6 can be rearranged as

$$\sigma_{xy} \simeq \frac{e^2}{h}\beta\sum_{n=1}^{d} F_n, \tag{9}$$

where we have also used the fact that $F_d = 0$. Thus, one can approximate the Hall conductivity by the sum of all Chern numbers of the indirect gaps (see Fig. S17). Note that for all $\beta > 0$ (i.e., decreasing occupation for increasing energy), the sign of the anomalous Hall conductance is purely determined by the sum of the Chern numbers of the indirect gaps and is robust to changes in filling factors produced by e.g. a shift in the Fermi energy.

---


\end{document}